\documentclass[conference,letterpaper,final]{IEEEtran}
\usepackage[utf8]{inputenc}
\usepackage{color}
\usepackage{textcomp}
\usepackage{url}
\usepackage{amsmath}
\usepackage{graphicx}
\usepackage{balance}
\usepackage{booktabs}
\usepackage{cite}
\usepackage[font=small]{caption}

\IEEEoverridecommandlockouts
\makeatletter
\let\@period=\@empty
\makeatother

\usepackage{array}
\newcolumntype{L}[1]{>{\raggedright\let\newline\\\arraybackslash\hspace{0pt}}m{#1}}
\newcolumntype{C}[1]{>{\centering\let\newline\\\arraybackslash\hspace{0pt}}m{#1}}
\newcolumntype{R}[1]{>{\raggedleft\let\newline\\\arraybackslash\hspace{0pt}}m{#1}}
\newcolumntype{P}[1]{>{\centering\arraybackslash}p{#1}}

\let\oldcaption = \caption
\renewcommand{\caption}[1]{\oldcaption{#1}\vspace{-5mm}}

\providecommand{\tabularnewline}{\\}

\makeatletter
\def\section{\@startsection{section}{1}{\z@}{1.5ex plus 0ex minus 0ex}%
{1sp}{\normalfont\normalsize\centering\scshape}}%
\def\subsection{\@startsection{subsection}{2}{\z@}{1.5ex plus 0ex minus 0ex}%
{1sp}{\normalfont\normalsize\itshape}}%
\def\subsubsection{\@startsection{subsubsection}{3}{\parindent}{0ex plus 0ex minus 0ex}%
{0ex}{\normalfont\normalsize\itshape}}%
\def\paragraph{\@startsection{paragraph}{4}{2\parindent}{0ex plus 0ex minus 0ex}%
{0ex}{\normalfont\normalsize\itshape}}%

\newcounter{protocol}[subsection]

\def\protocol{\@startsection{protocol}{3}%
{\z@}{1pt}{1pt}
{\normalfont\normalsize\itshape\bfseries}}
\makeatother

\begin{document}

\title{On Perception and Reality in\\ Wireless Air Traffic Communications Security}
\author{
\IEEEauthorblockN{Martin Strohmeier\IEEEauthorrefmark{1},
Matthias Schäfer\IEEEauthorrefmark{2},
Rui Pinheiro\IEEEauthorrefmark{3},
Vincent Lenders\IEEEauthorrefmark{4},
Ivan Martinovic\IEEEauthorrefmark{1}}\\[-0.3em]
\IEEEauthorblockA{
\begin{tabular}{C{0.23\textwidth}C{0.23\textwidth}C{0.23\textwidth}C{0.23\textwidth}}
 \IEEEauthorrefmark{1}University of Oxford, UK & %
 \IEEEauthorrefmark{2}TU Kaiserslautern, Germany & %
 \IEEEauthorrefmark{3}University of Hagen, Germany & %
 \IEEEauthorrefmark{4}armasuisse, Switzerland
\end{tabular}}
\vspace*{1em}
}
\maketitle
\begin{abstract}
More than a dozen wireless technologies are used by air traffic communication systems during different flight phases. From a conceptual perspective, all of them are insecure as security was never part of their design. Recent contributions from academic and hacking communities have exploited this inherent vulnerability to demonstrate attacks on some of these technologies. However, not all of these contributions have resonated widely within aviation circles. At the same time, the security community lacks certain aviation domain knowledge, preventing aviation authorities from giving credence to their findings.

In this survey, we aim to reconcile the view of the security community and the perspective of aviation professionals concerning the safety of air traffic communication technologies. To achieve this, we first provide a systematization of the applications of wireless technologies upon which civil aviation relies. Based on these applications, we comprehensively analyze vulnerabilities and existing attacks. We further survey the existing research on countermeasures and categorize it into approaches that are applicable in the short term and research of secure new technologies deployable in the long term.

Since not all of the required aviation knowledge is codified in academic publications, we additionally examine existing aviation standards and survey 242 international aviation experts. Besides their domain knowledge, we also analyze the awareness of members of the aviation community concerning the security of wireless systems and collect their expert  opinions on the potential impact of concrete attack scenarios using these technologies. 

\end{abstract}


\section{Introduction}\label{sub:Introduction}

Air traffic control (ATC) is the backbone of what is arguably the key means of personal transport in the modern world. As the traffic load continues to grow dramatically, ATC has to manage ever more aircraft. Large European airports, such as London Heathrow, experience spikes of more than 1,500 daily take-offs and landings, and industry forecasts predict that world-wide flight movements will double by 2030. Additionally, with the growing adoption of unmanned aerial vehicle (UAV) technology for civil applications, we can expect a further boost in air traffic in the coming years. 

Historically, ATC and its associated wireless communications technologies are rooted in the military. Most of the improvements in communication, navigation and surveillance (CNS) technologies are direct results of wartime developments \cite{spitzer2014digital}. For instance, surveillance radar systems and navigation functions which were developed originally for the armed forces were later adopted for civilian aviation. This change of purpose and application also shifted the threat models affecting these wireless technologies considerably. Where the military can often also rely on secrecy, security through obscurity, and superior proprietary technologies to prevail in an arms race, the requirements in a civil setting of worldwide collaboration are different. In this environment, a pure security by design approach, such as the protection of critical wireless communication through standard cryptographic countermeasures, would be highly preferable. Unfortunately, in the slow-changing industry of aviation, such a radical change in security approaches is not currently on the horizon (and arguably not easy to introduce at this point in time).

The civil aviation community emphasizes safety and has a sound and steadily improving safety record. Security, however, is not safety, and requires a different approach. While we encountered many helpful and interested people and institutions in aviation during our investigation, the prevalent feeling is still ``Why is security needed? Is air traffic communication not safe currently?''. Indeed, historically, few if any incidents have been recorded where communication technologies were maliciously exploited to cause distress to aircraft. Consequently, even recently developed aviation technologies that make the shift from traditional radar to modern digital communication networks do not include security by design in their specifications; instead the systems rely almost exclusively on redundancy.

However, with the widespread availability of cheap and powerful tools such as software-defined radios (SDR), the aviation community lost the considerable technical advantage protecting its communication in the past. This is illustrated by the recent proliferation of reports about potential cyberattacks on wireless ATC technologies. High-profile incidents, such as the case of hijacked emergency signals \cite{Bloomberg} or alleged military exercises causing aircraft to vanish from European radar screens \cite{Reuters}, created a lot of speculation in the media about the potential impact of insecure technologies on the safety of air traffic \cite{CNN,Wired}. Disregarding the accuracy of individual reports, these speculations are directly caused by the fact that such attacks are potentially feasible, which has been proven recently by hackers \cite{Costin,bradhaines2012} and the academic community \cite{Schaefer13}. Much further research has been conducted on the security of newly-developed ATC protocols, and the digital avionics systems installed in modern aircraft \cite{Sampigethaya,teso2013aircraft,lundberg2014security}.

Meanwhile, following these revelations, accusations of overblown media reporting of cybersecurity issues have been made by members of the aviation community. Some claim the impossibility of the cited hacks of aircraft IT systems in the real world \cite{Register,defcontruth}, or doubt the impact of attacks on wireless air traffic communication systems in practice due to the widely deployed checks and balances common to aviation \cite{NATS}.

These instances show that, unfortunately, many who understand wireless security, do not have appropriate aviation expertise. Likewise, many stakeholders in aviation know the processes and procedures but do not realize the severity of modern cybersecurity issues. Our work aims to integrate these colliding perspectives in a realistic model and inform future discussion about wireless security in aviation.

There is much existing work in the wider area of aviation cyber security; in Section \ref{sub:countermeasures} we survey those articles that relate to the wireless communications technologies on which we focus in the present work. For an introduction to the topic, the reader is referred to Sampigethaya et al., who focus on future ``e-enabled'' aircraft communications and their security \cite{Sampigethaya} and highlight the challenges and problems of these modern cyber-physical systems \cite{sampigethaya2013aviation}.  Likewise, \cite{sampigethaya2008secure} surveyed the security of future on-board and off-board avionics systems, relating not only to wireless ATC communication but to the electronic distribution of software and air health management.

\subsection*{Contributions}
The goal of this paper is to consolidate academic and hacker security knowledge, information from aviation technology standards, and expert opinions from the aviation domain, both detailed and aggregated. We systematize the collected knowledge along three dimensions:

\begin{itemize}
\item We provide a systematization of all relevant wireless technologies considering their applications in aviation: air traffic control, information services, and navigation aids.  We explain their usage within the aviation system, detail their technical features, and discuss their safety impact.

\item We systematically integrate knowledge from the academic and hacker communities with technology standards to analyze the security of these applications, providing a complete overview over their vulnerabilities, existing attacks, and countermeasures. We further categorize all proposed countermeasures according to their applicability time frame. 

\item As not all of the required aviation knowledge is currently codified publicly, we additionally survey 242 international aviation experts to capitalize on their domain knowledge.  We further examine the awareness of the aviation community concerning wireless security and collect expert opinions on the potential safety impact of attacks on these technologies. 

\end{itemize}


Our systematization identifies discrepancies between stakeholders and provides an integrated view of aviation communications. To the best of our knowledge, this is the first investigation of this scale, including knowledge from more than 200 participants from different aviation groups, over 400 comments on the topic of cyberattacks, a practical assessment of safety impact and security features of all relevant technologies, and an evaluation of nine concrete hypothetical attack scenarios.

We highlight the urgent need to increase knowledge and awareness of the discussed problems within both communities in order to improve short- and long-term wireless air traffic security. In particular, we argue that without awareness among professionals and entities in aviation of the criticality of existing vulnerabilities, the necessary change will likely not come about before a fatal real-world accident occurs.

The remainder of this work is organized as follows: Section \ref{sub:Overview} gives an overview of air traffic communication and our adversarial model. Section \ref{sub:survey} provides the background to our survey. Section \ref{sub:protocols} analyzes the technologies of ATC, information services, and navigation aids, including their security. In Section \ref{sub:Impact}, we discuss the aviation experts' evaluation of concrete attack scenarios. Section \ref{sub:countermeasures} categorizes the existing countermeasures and identifies current directions in aviation security research. Section \ref{sub:Conclusion} concludes this work.

\begin{table}
\scriptsize
\begin{centering}
\begin{tabular}{|>{\raggedright}p{1.1cm}|>{\raggedright}p{6.5cm}|}
\hline 
\textbf{Abb}. & \textbf{Full Name}\tabularnewline
\hline 
\hline 
\multicolumn{2}{|l|}{\textbf{Air Traffic Control}} \\
\hline 
VHF & Voice (Very High Frequency)\tabularnewline
\hline 
PSR & Primary Surveillance Radar\tabularnewline
\hline 
SSR & Secondary Surveillance Radar (Mode A/C/S)\tabularnewline
\hline 
ADS-B & Automatic Dependent Surveillance-Broadcast\tabularnewline
\hline 
CPDLC & Controller–Pilot Data Link Communications\tabularnewline
\hline 
MLAT & Multilateration\tabularnewline
\hline 
\multicolumn{2}{|l|}{\textbf{Information Services}} \\
\hline 
ACARS & Aircraft Communications Addressing and Reporting System\tabularnewline
\hline 
TCAS & Traffic Alert and Collision Avoidance System\tabularnewline
\hline 
FIS-B & Flight Information System-Broadcast\tabularnewline
\hline 
TIS-B & Traffic Information System-Broadcast\tabularnewline
\hline 
\multicolumn{2}{|l|}{\textbf{Navigational Aids}} \\
\hline 
GPS & Global Positioning System\tabularnewline
\hline 
VOR & VHF Omnidirectional Radio Range \tabularnewline
\hline 
ILS & Instrument Landing System\tabularnewline
\hline 
NDB & Non-directional Beacon\tabularnewline
\hline 
DME & Distance-measuring Equipment\tabularnewline
\hline 
\end{tabular}
\par\end{centering}

\protect\caption{Short-handles and full names of aviation communication technologies, systematized into applications. \label{tab:shorthandles} }
\end{table}%

\section{On the Problem of Air Traffic Communications Security}\label{sub:Overview}

One goal of this paper is to show the obvious mismatch between security research and the aviation community concerning their approaches to the problem of air traffic communications security. We consider the hypotheses that either existing wireless security flaws are widely known within aviation but are not deemed realistically exploitable, or that awareness in the community simply does not exist. 

\begin{figure*}[t]
\centering
\includegraphics[bb=2bp 0bp 790bp 465bp,clip,width=0.91\textwidth, height=240bp]{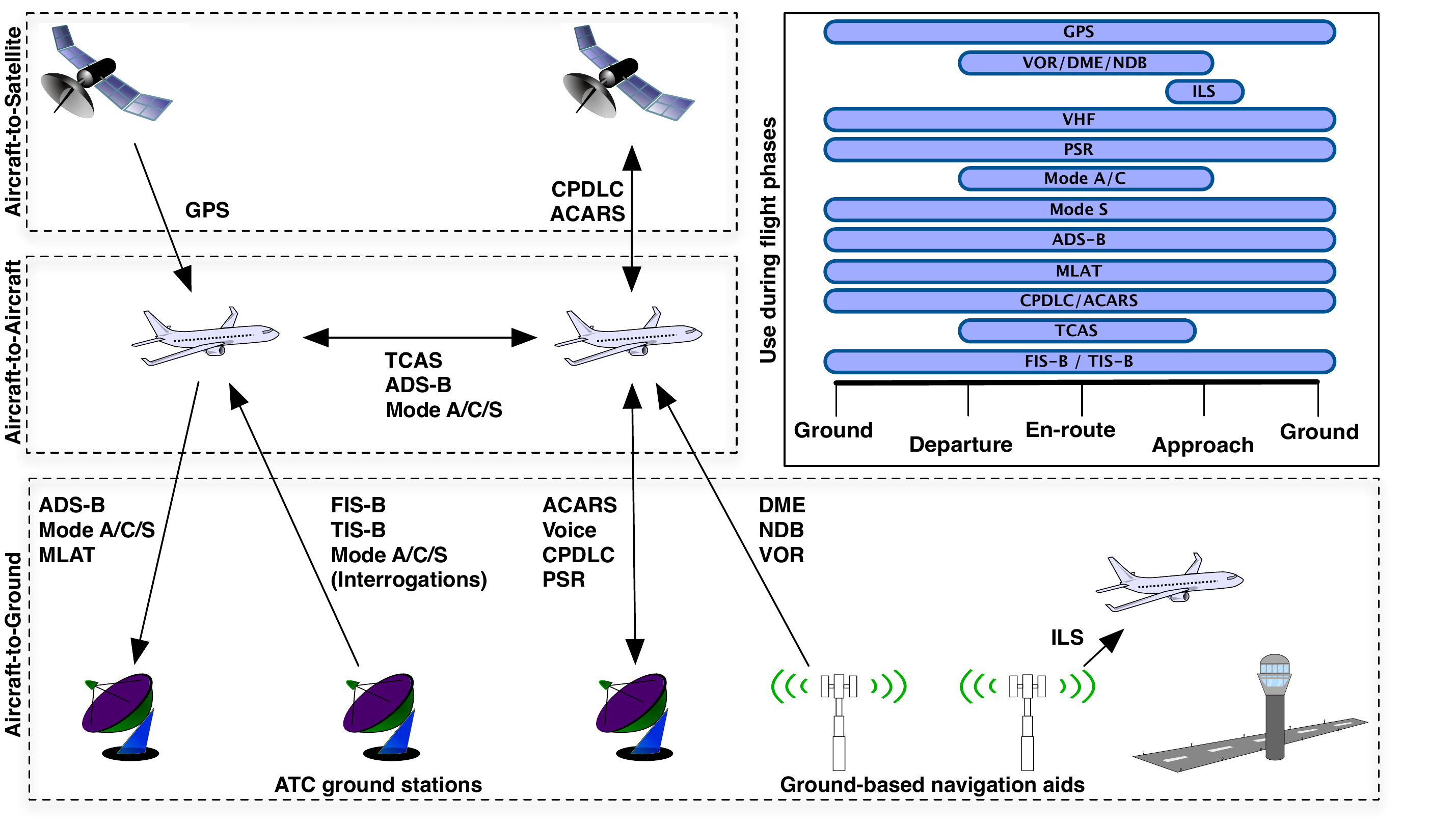}
\caption{An overview of the wireless technologies used in air traffic communication, between ground stations, aircraft and satellites. The arrows indicate the direction of the communication for each protocol. The Gantt chart shows their typical usage during the different flight phases.}
\label{fig:ATC-overview}
\end{figure*}

As our investigation shows, aviation professionals at large are unaware of the state of security in aviation technology. A majority of participants falsely believe, for example, that even the most common surveillance technologies offer authentication, clearly contrasting with the security community's knowledge. Indeed, our findings are in line with research from cognitive science, which suggests that so-called ``expert blind spots'' exist: i.e., having a large amount of domain-specific knowledge may prove disadvantageous on tasks such as forming remote associations among disparate concepts \cite{nathan2001expert}.

We believe it is crucial to challenge the status quo of cybersecurity in aviation, and to integrate it with the reality of insecure technologies. Naturally, changes depend on increased awareness from the users of these technologies, who can then demand and help to develop solutions. On the other hand, legacy requirements and the unique aviation environment often prevent the straightforward use of established security solutions such as cryptography in the short and medium term. Still, with the help of the security community, it could be feasible to develop systems that improve communications security in aviation within a shorter time span than the typical development cycles of 20-30 years. We discuss current research trends and challenges associated with practical solutions in Section \ref{sub:countermeasures}.


In the remainder of this section, we give a brief initial overview of the communication systems analyzed later in this paper, and discuss the main factors that are considered as mitigating potential attacks within the aviation community.

\subsection{Overview of Air Traffic Communications}
In this paper, we focus on the whole picture of aviation as found under Instrument Flight Rules (IFR), where navigation depends on electronic signals. IFR usually apply to large commercial aircraft, even though they are open to any aircraft with the necessary equipment, including general aviation (i.e., civil aviation that is not a scheduled air service or air transport for remuneration or hire). However, many of our findings also apply to flying under Visual Flight Rules (VFR). Under European Organisation for the Safety of Air Navigation (EUROCONTROL) and Federal Aviation Administration (FAA) regulations, aircraft under VFR need to equip fewer communication systems and enjoy more freedom in their choice compared to  commercial aircraft, but also have fewer options in case of their failure.

Fig.\,\ref{fig:ATC-overview} provides a comprehensive, high-level picture of currently employed wireless communication technologies in commercial aviation, focusing on the interactions between the technologies and their utilization during different flight phases. To aid the reader's understanding throughout the paper, we have collected the most important acronyms in Table \ref{tab:shorthandles}. We generally use broad definitions of \textit{communications} and \textit{protocols} which include analog technologies as well as message-based protocols transmitting digital data.

In order to focus on the systems view, we have divided all technologies into three categories according to their application: air traffic control, information services, and navigation aids. ATC protocols are used to enable communication between controllers and pilots or their aircraft. They include VHF, PSR, SSR, ADS-B, MLAT, and CPDLC, which are used during all flight phases, usually on line-of-sight frequencies, although the use of satellite communication is possible (e.g., CPDLC). Information services offer a more general platform for the exchange of data such as weather and traffic information: we discuss ACARS, TCAS, FIS-B, and TIS-B. Navigational aids, namely GPS, VOR, DME, NDB, and ILS, locate the aircraft's position in space and aid in the approach. Navigation aids receive signals from separate infrastructure on the ground or, in the case of GPS, from satellites.

While there is considerable fragmentation among aviation systems all over the world, we aim to be as comprehensive and generally applicable as possible. Technologies and procedures related to military systems, both secret in nature and exclusive to a country's air force, are out of the scope of this paper. We further appreciate that even some of the same systems differ across regions, and due to space limitations, we cannot address every exception. However, our findings are broadly applicable, as the underlying technologies and principles are the same.

\subsection*{Air Traffic Management}
Another way to categorize these technologies is to embed them in the hierarchical model of air traffic management (ATM) and to look at their time horizons and feedback loops. After Klein \cite{kleinatmhierarchy}, modern ATM can be divided into long-term, centralized, strategic flow planning on the (supra-)national stage; medium-term flow planning and control of smaller sectors; and finally, the highly decentralized, immediate-term, aircraft guidance and navigation (see also \cite{zhang2012hierarchical}).

As the time horizons narrows, ATM moves from the long-term objective of safe and efficient flow management using filed flight plans and historical movement data towards the ultimate goal of accident prevention. Each control layer has different deadlines depending on the operational requirements. For example, long-term flow planning requires historical flight data collected through ATC technologies but their integrity does not provide an immediate concern for safety. In this work, we focus on the short and medium term, where an attack on wireless ATM technologies has the largest safety impact. This comprises technologies transmitting data that is relevant for hours (e.g., weather) or highly critical systems for collision avoidance, which operate in the matter of seconds.

\subsection*{Potential Future Technologies}
The International Civil Aviation Organization (ICAO), EUROCONTROL, and the FAA have started planning for further upgrades of the current communications systems and are seeking to develop new data links. Specifically, L-band Digital Aeronautical Communications System (L-DACS) and Aeronautical Mobile Airport Communications System (AeroMACS) are supposed to replace the current VHF system. Since these systems can provide much higher data throughput comparing to the existing data links, some of the applications currently provided by other technologies could also utilize these new technologies one day. Thankfully, L-DACS and AeroMACS have begun to at least consider the issue of wireless security and some corresponding designs are already included by the specifications or will be in the future.

Unfortunately, L-DACS is still in the very early specification phase and in line with typical technological cycles in aviation will not be deployed before the 2030s \cite{iacognap}. Furthermore, since its specification is not finished -- many parts are up in the air and there are still competing proposals -- they could strongly benefit from an immediately increased awareness about security concerns in aviation, which we aim to provide with our work.

AeroMACS takes the form of a profile of IEEE 802.16-2009 \cite{ieee2009wimax}, known as WiMAX. It intends to provide a surface data link for use at the airport, allowing ATC, airlines, and airports to communicate with the aircraft \cite{aeromacseurocontrol}. It has line-of-sight range of up to 3\,km per cell and uses commodity radios to communicate. While the current standards include cryptography, making it a serious step forward, AeroMACS will not solve the security problems currently found in aviation. Besides the prevalent issue of long deployment time frames (the beginning of deployment is not projected before the middle of the next decade), many security questions such as the protection of management frames are still undecided \cite{aeromacseurocontrolsecurity}. Most importantly, AeroMACS will only be able to replace current data links on the ground and in the immediate vicinity of an airport, leaving the vast amount of air traffic communication unprotected.

AeroMACS is further along in the development cycle compared to L-DACS, with test deployments going on at some airports around the world. However, at the time of writing, many of the necessary avionics standards and specifications were still in the planning phase \cite{aeromacseurocontrol}. Thus, it was not possible to include it in our survey but, like L-DACS, it should see strong input from the security community as soon as possible.





\subsection{Threat Model}
In this section, we define the threat model that wireless systems in aviation face. We distinguish between the traditional adversarial model and the recently emerged model based on a) the widespread distribution of accessible software-defined radios and b) the ongoing move from analog to digital communication systems (as pointed out in, e.g., \cite{mahmoud2014aeronautical}). We consider \textit{active} adversaries with the capability to eavesdrop, modify, and inject data on the communications channel.

\vspace{10pt}

\subsubsection{Traditional Aviation Threat Model}
The traditional threat model has been implicitly and explicitly used in aviation since the introduction of radio communication and radar in civil air traffic control in the first half of the 20th century. Surveillance radar, navigation, and communication systems originated from military applications and were later integrated into the civil aviation airspace \cite{Swords86}. We characterize the threat model used as comparatively naive, reflecting the general state of computer security considerations within industrial and infrastructure systems during this period. In short, the model makes the following main assumptions on which today's aviation communications security is still based \cite{strohmeier2016cycon}:

\vspace{5pt}

\begin{itemize}
\item \textbf{Inferior technological capabilities:} Active adversarial capabilities were subscribed only to military and nation-state attackers with the ability to conduct electronic warfare \cite{adamy2001ew}.
\item \textbf{Inferior financial capabilities:} Similarly, it is further assumed that electronic devices capable of distorting radar are financially out of reach for all but the most capable attackers.
\item \textbf{Requirement of inside knowledge:} An impactful attacker needs to be an \textit{insider} to obtain the necessary knowledge of communication systems and general aviation conduct.
\item \textbf{Use of analog communication:} Typical attacks on analog communications are easier to detect for the \textit{user}, e.g. somebody hijacking the voice channel or causing a denial of service at the PSR will typically be detected immediately.
\end{itemize}

\vspace{10pt}

\subsubsection{Modern Threat Model}
With the technological advancements of the late 1990s and 2000s, the threat model changed drastically, as the assumptions about adversaries ceased to hold. For instance, in the 1990s, SDRs were first practically adapted for military and closed commercial use \cite{dowla2003handbook}. However, open-source projects such as GNURadio \cite{blossom2004gnu} released in 2001, and finally the availability of cheap commercial off-the-shelf (COTS) software-defined radio transceivers spread adversarial capabilities to a large and expanding group of people. They enable a broad community with basic technological understanding to receive and process, craft and transmit arbitrary signals -- including those used in aviation. Contrary to the pre-SDR era, hardware need not be purpose-built anymore (requiring considerable technical and financial resources) but can simply be programmed and re-programmed on the fly with the necessary code and knowledge easily shared via Internet.

We make the following assumptions for an adversary model that is adequate for wireless communications security in modern aviation and which we use throughout this work:

\vspace{5pt}

\begin{itemize}
\item \textbf{Increased digitization and automation:} As we will point out throughout the paper, there is a general trend in aviation towards transmitting data such as flight clearances with unauthenticated digital communication networks. While attacks on analog technologies such as VHF have been included in the traditional threat model, new \textit{digital attacks} are emerging which are easy to execute, potentially devastating, and difficult to detect on the data link level for increasingly \textit{automated systems} and their users \cite{wolf2014information}.
\item \textbf{Increased technological capabilities:} With the widespread availability of cheap SDR technology, it is reasonable to assume that a large amount of people are capable of conducting wireless attacks on aviation systems. The financial barrier is all but gone with SDR receivers available from as little as \$10 while capable senders cost less than \$100 with a strong downwards trajectory. In conjunction with downloadable software, this development enables a new class of \textit{unsophisticated} attackers.
\item \textbf{Easy availability of aviation knowledge: } Attackers have general knowledge about processes and conventions in aviation communications. Syntax and semantics of wireless protocols can be obtained by \textit{outsiders} through openly accessible means, such as specification protocols, forums, planespotting websites, and finally by capturing and examining real-world communication data. 
 
\end{itemize}

In summary, attack capabilities shifting from military adversaries to script kiddies, hobbyists, white hat hackers, cybercrime organisations, and terrorists increase the likelihood of attacks manifold. In conjunction with the move towards unsecured digital networks, and increased deployment of homogeneous COTS hard- and software, the new threat model goes beyond denial of service through traditional jamming and requires us to rethink and address wireless security in aviation. 

\subsection{Mitigating Factors}
Before our in-depth analysis, we want to consider some of the existing factors that can help mitigate the threat of wireless attacks on air traffic applications. These factors can be divided into a) \textit{rules and procedures} and b) the existence of \textit{redundant communication systems}. Both factors do not traditionally consider security, i.e., malicious attacks on the analyzed systems, but focus on maximizing the overall safety of the aircraft, regardless of the interfering elements.

\subsubsection{Procedures}
Procedures and practices for systems failures are plentiful in aviation and try to cover all imaginable cases. While they are purely aimed at non-deliberate failures, many of them (e.g., lost communications procedures, the FAA's Emergency Security Control of Air Traffic (ESCAT) plans for major crises such as 9/11, ATC Zero procedures to handle local failures of ATC centers) will have mitigating consequences for deliberate attacks as well. Overall, human factors are of crucial importance in aviation. Many pilots have experienced incorrect instrument readings and are trained to check and double-check at all times. On the ATC side, if a controller noticed more than one aircraft using the same transponder identity, they would call the plane and provide a different one, followed by a request to set an IDENT flag that will be displayed on the ATC screens.

Unfortunately, even without malicious attackers, procedural defenses are not always successful. One particular example is when priorities and instructions of different protocols - both human and technical - are not well-defined or even conflicting, such as in the fatal 2002 Überlingen mid-air collision.\footnote{The two aircraft had received different instructions from ATC and their on-board equipment. By each following a different procedure, their collision course was not resolved. See, e.g., \url{http://goo.gl/WZFceZ} for a full analysis.} 

On top of this, it is highly unlikely in reality that \textit{every} single malicious communications interference is detected by humans and defended using only rules and procedures. Sophisticated attacks on communication are simply never considered in aviation training. Existing procedures (e.g., FAA regulations and the Aeronautical Information Manual \cite{2015faraim} for general aviation in the US) cover only faulty systems, assuming that available information is genuine and has not been maliciously altered. Furthermore, any new generation of pilots is trained to rely on instruments and digital systems even more,\footnote{This reliance is illustrated by some recent incidents, most notably the Air France disaster where the pilots flying trusted instruments even though they knew they were unreliable \cite{et2012final}.} motivating the increased need to secure their integrity.

\subsubsection{Redundancy}
Related to procedure-based mitigations is the concept of redundancy. The most common type is the use of \textit{hardware redundancy} to improve the availability of a system. This can, for example, include numerous duplicate senders or receivers within the aircraft, or completely independent systems. Similarly, any ATC ground station utilizes several receivers for their surveillance or multilateration systems, making the failure of one, or even a few, unproblematic. 

Besides this, there are procedures for wireless systems failures that rely on different systems for redundancy. When a primary data source is not available, there are often other systems at the disposal of a pilot or a controller to obtain the required information. For example, when a pilot under instrument flight rules loses ILS capabilities, they can use other navigation aids (e.g., DME/VOR) to aid the approach.

On the ATC side, an aircraft with a Mode A/C/S transponder failure may still be tracked using PSR. However, the altitude and ID, among other information, are lost, causing a significant drag on the controller's awareness and attention while maintaining the required separation.  

However, this type of \textit{technological redundancy} has inherent security flaws. As the technologies were not developed for redundancy, but simply happened to become alternative options because of legacy reasons, relying on older technologies comes with a degradation of content quality. For example, separation minima may need to be increased when SSR systems fail, or an aircraft without working ILS receiver may lose its ability to land on some airports in bad weather conditions. On top of this, the availability of systems can differ across airports, making general assurances difficult. Overall, technological redundancy is effective in many cases when it comes to preserving safety. However, it fails when there is no suspicion of malicious activity by the user(s), or there are attacks on the procedures themselves (i.e., multiple technologies are targeted). 
\vspace{5pt}
\section{Survey on Cybersecurity in Aviation}\label{sub:survey}

To collect additional domain knowledge, data on the awareness of cybersecurity in aviation and quantify the impact of attacks, we conducted a survey across all aviation circles. This survey is the first to address these issues publicly, and we are thankful to aviation authorities and air navigation service providers (ANSP) for their help. Our survey was conducted fully anonymously over the internet using SurveyMonkey, with recruiting done via mailing lists of ANSP, airlines and other aviation-related organisations, as well as two closely-modera\-ted forums for pilots. We did not aim to survey people with knowledge in computer security, but to obtain a realistic opinion of the aviation community as a whole. 

The three main research questions that the survey looks to answer are a) Which technologies are considered to have the biggest impact on safety? b) Are aviation stakeholders aware of security issues in the wireless technologies they utilize? c) If yes, are these issues considered a concern towards safety? 

We analyze the answers to these questions after discussing the design of the study, and the demographics of the respondents. We then discuss the respondents' assessment of nine concrete hypothetical attack scenarios in Section~\ref{sub:Impact}.  


\subsection{Survey Limitations} 

We planned and conducted our survey with the help of private pilots and a full-time professional air traffic controller among the authors. They advised us on the appropriate question language, and provided us with the necessary aviation expertise and background at every stage during the design, implementation, and execution of this survey. It is worth noting that a survey-based analysis is an accepted tool in aviation research. For example, the authors in \cite{silva2015safety} recently used it to analyze the safety of the FIS-B protocol. 
 
However, we are aware of the limitations of our approach. We tried to mitigate confounding factors through our design, but we acknowledge some potential limitations caused by the characteristics found in aviation technology:\\

\begin{itemize}
\item \textbf{Proprietary systems:} Typically, systems are implemented by different companies following loose standards. Even some protocols (e.g., ACARS) have proprietary elements not freely available. To counteract this problem, we abstracted away from the concrete implementations. We designed the questions such that we could draw more general conclusions on the respondents' knowledge of the systems' security.
\item \textbf{Fragmentation:} Likewise, there is a forest of different systems, regulations, and processes in aviation. Depending on the airspace, the availability, knowledge, and usage of the discussed protocols differ. However, we mitigated this problem by surveying experts from many countries, making sure their judgment of security in aviation technologies did not vary significantly.
\item \textbf{Representativeness:}
Considering distribution and potential self-selection, we do not claim that our results are necessarily representative of the aviation community. Yet, when reconciling with comments and conversations with experts, we believe in their validity.
\end{itemize}

Overall, we believe it is an important task to abstract away from single technologies and gather a more systemic picture of the awareness on wireless security in aviation as a whole. 

\subsection{Ethical Considerations}
We acknowledge the potentially sensitive nature of analyzing air traffic communication protocols and raising awareness of easily exploitable security flaws. ATC is considered a critical public infrastructure protecting the lives of billions of people every year but while we argue that wireless attacks in general become more and more feasible for non-state actors, some isolated protocols have been widely analyzed by both hacker communities and academic researchers. The vulnerabilities inherent in these protocols have been known for years and even concrete exploits are widely available in hard- and software. While we consider the overall state of aviation communications security a serious one, redundancy and existing processes still mostly protect modern airspaces currently. 

We conducted our survey fully anonymously, to protect respondents from potential repercussions when speaking about the security of ATC systems or disclosing safety problems. We follow a responsible disclosure process, working with ATC institutions during the planning and discussions phase of our research. We notified these institutions of our results and plan to work closely with them in the future. We obtained ethical approval from the University of Oxford's Social Sciences \& Humanities Inter-Divisional Research Ethics Committee (IDREC) under the Ref No: SSD/CUREC1A/15-033. 

\begin{table}[t]
\small
\begin{tabular}{|>{\centering}m{3cm}|>{\centering}m{2.55cm}|>{\centering}m{1.82cm}|} 
\hline 
\textbf{Group} & \textbf{\# Respondents} & \textbf{Share}
\tabularnewline
\hline 
\hline 

Private Pilot & 77 & 32.0\% 
\tabularnewline
\hline 

Commercial Pilot & 59 & 24.5\% 
\tabularnewline
\hline 

Civil ATC & 37 & 15.4\% 
\tabularnewline
\hline 

Aviation Engineer & 10 & 4.1\% 
\tabularnewline
\hline 

Aviation Authority & 7 & 2.9\% 
\tabularnewline
\hline 

Military Pilot & 5 & 2.1\% 
\tabularnewline
\hline 

Military ATC & 4 & 1.7\% 
\tabularnewline
\hline 

Other & 42 & 17.4\% 
\tabularnewline
\hline 

\end{tabular}\protect\caption{Occupations of survey respondents ($n=242$). } \label{WorkTable}
\end{table}

\begin{figure*}[t]
\centering
\includegraphics[bb=12bp 0bp 1260bp 440bp,clip,width=0.935\textwidth, height=150bp]{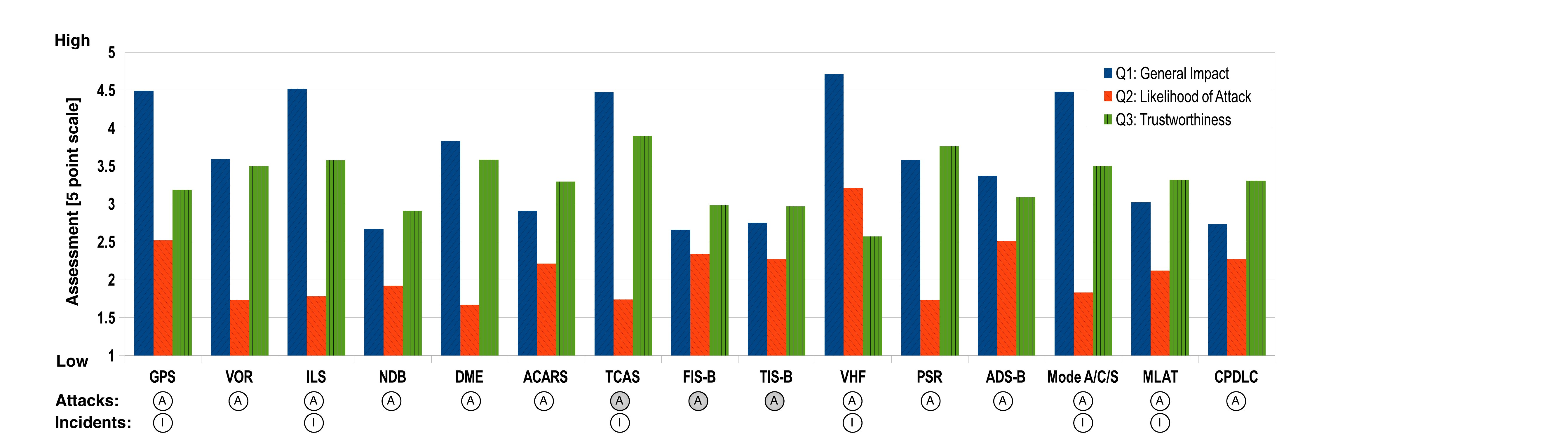}
\caption{Assessment of 1) the flight safety impact, 2) the likelihood of being attack targets and 3) the trustworthiness against manipulation of each protocol. The respondents ($n=235$) answered three questions. Q1: ``\textit{How would you rate the flight safety impact of each of these technologies?}'', Q2: ``\textit{How do you rate the likelihood that a malicious party injects false information into these technologies?}'', and Q3: ``\textit{How would you rate the trustworthiness of information derived from these technologies against intentional manipulation by a malicious party?}''. Answers were provided on an equidistant 5-point-Likert scale. The circles show the availability of published attacks and (publicly or privately) reported incidents for each protocol. Gray circles are protocols vulnerable by extension as they depend on data from other vulnerable protocols.}
\label{fig:survey-overall}
\end{figure*}

\subsection{Initial Survey Results}

\subsubsection*{Demographics}
We had 242 completed surveys, 110 or 45.5\% from a controlled dissemination (CD) using internal mailing lists of aviation-related companies and authorities, and 132 or 54.5\% from a separate open dissemination (OD) on internet aviation forums. We compared the results of both and found no significant differences in the respondents' evaluations apart from their professions: 55.7\% of OD respondents were part of general aviation (GA), i.e., private pilots and not otherwise working in aviation, compared to only 3.6\% of CD respondents. We analyze the responses as a whole unless stated otherwise. Not all questions have been answered by all experts, accounting for differences in the number of respondents as indicated for all reported results.

The participants' aviation experience was fairly evenly distributed, with 32\% having 20 years or more, and about 22\% offering an expertise of less than 5 years, 5-10 years, and 10-20 years, respectively. The top working countries were the UK (37.7\%) and the US (23.3\%). A further 37.3\% work in Continental Europe, with 4 respondents from other countries around the world (Indonesia, Hong Kong, Canada, UAE).

As illustrated in Table \ref{WorkTable}, most of our respondents were private (32\%) or commercial pilots (24.5\%) followed by civil air traffic controllers (15.4\%) and aviation engineers (4.1\%). We had responses from professions as varied as Air Traffic Safety Electronics Personnel, researchers at ANSPs, ATC technicians, aviation software developers, or Flight Information Service Officers/ Instructors amongst others. The average response time length was 31 minutes and 10 seconds, which, along with the plethora of comments, shows that the respondents took the survey seriously.

\subsubsection*{Self-assessed Knowledge and Work Environment}
The respondents judged their air traffic communication knowledge above average for their field (3.76 out of a symmetric, equidistant 5-point Likert scale, where 1 is ``very bad'' and 5 is ``very good''). The technologies that the respondents considered themselves most familiar with are GPS, VHF, ILS, VOR and Mode A/C/S. This is explained by the prevalence and importance of these technologies in current aviation processes: Mode A/C/S and VHF are also the most relied upon technologies, followed by TCAS, GPS, and ILS. 

\subsubsection*{Impact Assessment}
While our detailed analysis follows in later sections, Fig.\,\ref{fig:survey-overall} shows that VHF, ILS, GPS, Mode A/C/S, and TCAS are considered the technologies with the highest safety impact, all around 4.5 out of 5 (``very high''). But even the least important technologies (which are either outdated -- NDB and VOR -- or not widely operational yet -- CPDLC, FIS-B, TIS-B) still have a moderate impact according to the respondents. 
Generally, the isolated impact of a single protocol is often considered limited. A German controller represents the feelings of many (but not all): ``\textit{in case of loss there are still backup systems and cross-check possibilities so there are no really high impact ratings}'', illustrating the traditional safety approach of redundancy extended to security.



\subsubsection*{Security Assessment}
As shown by the red (middle) bars in Fig.\,\ref{fig:survey-overall}, the navigation aids (with the exception of GPS), and TCAS are considered by far the most trustworthy protocols when it comes to the likelihood of manipulation by an adversary. VHF is considered the least trustworthy ($>$60\% say it is ``very likely'' or ``likely'' to be attacked) by the participants. Of the digital protocols, ADS-B and GPS were considered the most likely to be attacked. We speculate this might be due to the raised awareness caused by widely publicized attacks. Still, all technologies except VHF are considered relatively unlikely to be attacked, despite the prevalence of reported attacks and incidents indicated in Fig.\,\ref{fig:survey-overall} and discussed in detail later.

Consistently, the technology most respondents believed lacked integrity and authentication mechanisms is VHF (see Fig.\,\ref{fig:integrity-impact-atc}), followed by NDB. For all other technologies, the respondents believed they offered such security mechanisms, or did not know. In general, there was a high uncertainty ($>$15\% of respondents), even for the most well-known technologies. 
\section{Wireless Communication Technologies in Aviation}\label{sub:protocols}

In this section, we analyze the security features and impact of all considered air traffic communication technologies more closely, while contrasting it with the opinions of the surveyed aviation experts. As the application of the technologies determines the consequences and severity of security breaches, we divide the technologies accordingly in three categories. In principle, all technologies in a category can at least partly serve the required application. 

\emph{Air traffic control} comprises technologies which support air traffic services. This includes communication links between controllers and pilots and technologies for monitoring air traffic. \emph{Information services} are technologies which provide information to pilots to improve their situational awareness (e.g., weather or traffic information). Finally, \emph{radio navigation aids} provide the pilots means to navigate in the air during different phases of flight.

\subsection{Air Traffic Control}\label{sub:AirTrafficControl}
\begin{figure*}[t]
\centering
\includegraphics[bb=14bp 2bp 390bp 285bp,clip,width=0.446\textwidth,height=130bp]{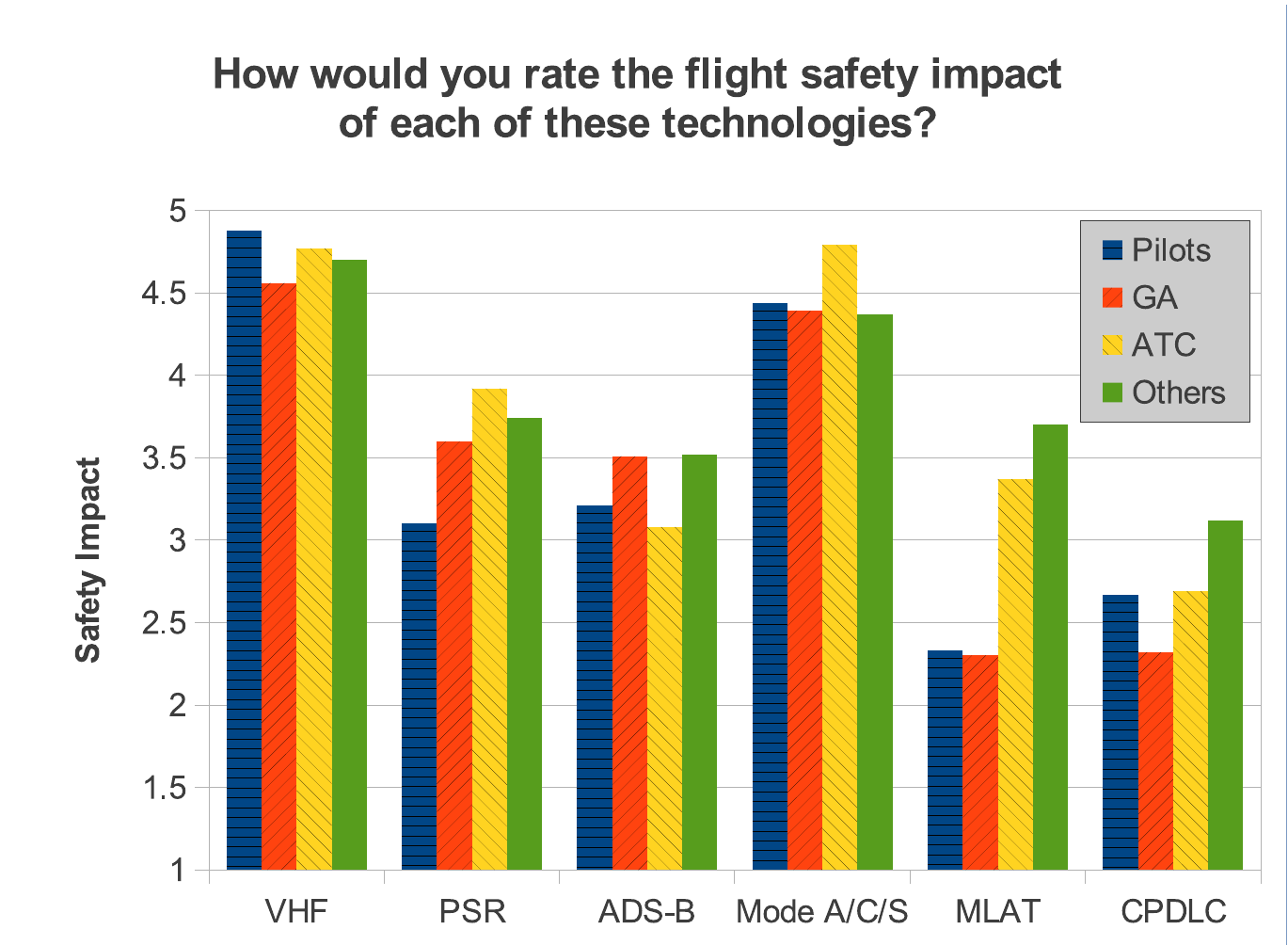}\includegraphics[bb=23bp 10bp 555bp 360bp,clip,width=0.476\textwidth,height=135bp]{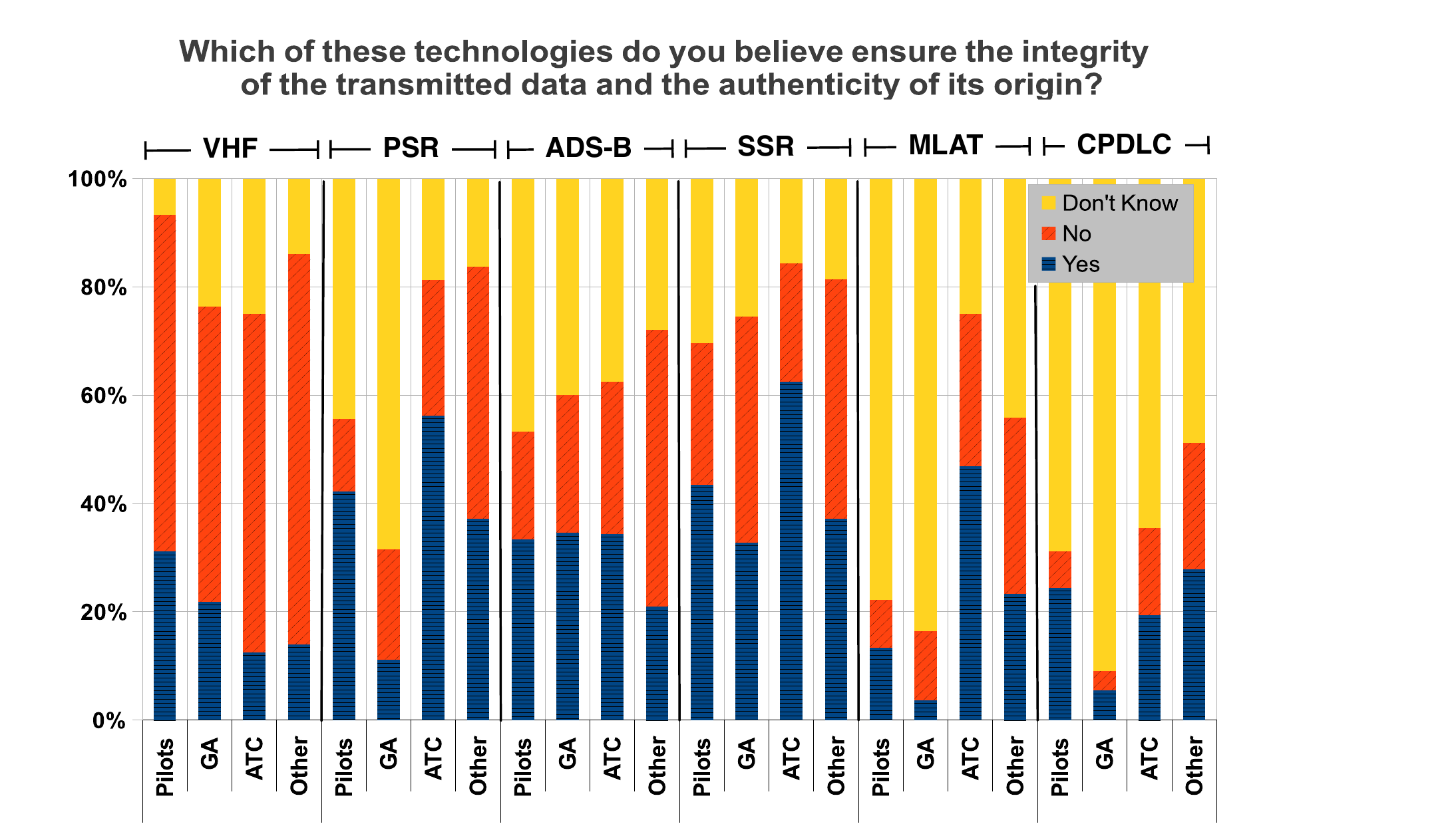}
\caption{Assessment of ATC technologies' safety impact and security capabilities (authentication and integrity). Data gathered from 43 commercial pilots, 55 private pilots, 32 controllers, 45 others.}
\label{fig:integrity-impact-atc}
\end{figure*}

ATC protocols enable communication between controllers and pilots or their aircraft. They establish information about the aircarft's position and intent and thus the safety of the airspace. Table \ref{fig:ATC-table} provides an overview of their technical details. Fig.\,\ref{fig:integrity-impact-atc} breaks down two aspects of the protocols based on survey data: impact according to stakeholders, and whether respondents believe that they offer integrity and authenticity.

\protocol*{Voice (VHF)}
Voice communication \cite{Anx10V3} is the primary means of communication between ATC and the aircraft which is reflected by its high safety impact rating across all surveyed groups in Fig.\,\ref{fig:integrity-impact-atc}. It is used to transmit all ATC instructions (clearances) to the aircraft, which are acknowledged by the pilot, as well as pilots' reports and requests to ATC. Flight information service, weather, and airport information broadcasts can also be provided by voice communication. It is further used for operational communication between the airline operator and the aircraft, as far as the aircraft is in range of the operator's transmitter. Voice communication is conducted by analog radio on VHF and HF (outside VHF range, e.g., over oceans) \cite{Anx10V5}.

\subsubsection*{Security Considerations}
Successful VHF communication depends on the correct understanding of the message by the communication partners. Besides taking care of human factors, a high quality signal must be ensured. Simultaneous use of the frequency leads to a partial or full denial of service (DoS) in practice. Despite the fact that VHF employs amplitude modulation (AM), which allows reception of multiple channels on the same frequency, it is difficult to maintain service with an attacker dedicated to disturb the intended communication.
Authentication procedures are available for military flights only, in case the pilot insists. As they are time and capacity consuming, they are not applied for civil flights.
When CPDLC is not available, as in most regions currently, or the aircraft is not equipped with it, there is no backup protocol for VHF. Losing this main communication layer within highly-populated airspaces results in a severe threat. 

VHF is the least trusted protocol in our survey by a significant margin (see Fig.\,\ref{fig:survey-overall} and \ref{fig:integrity-impact-atc}). Many experts report actual experience with non-legitimate uses of the frequency, one noting ``\textit{VHF is an increasingly common comms signal to be maliciously emulated by non-involved parties. Particularly on tower frequencies. Anyone can buy an aviation transceiver without licence.}'' while others mention problems with pirate radio stations. There are some related incidents with spoofed voice communication reported in the aviation literature and on the Internet \cite{stelkens2015towards}, and recent works also discuss the urgent need to improve the security of VHF \cite{fantacci2009secure,stelkens2015towards}. While intruders are quickly detected through changes in signal or voice levels, between 30\% and 40\% of VHF users believe it offers technical integrity and authenticity checks. Furthermore, attackers could effectively disable VHF (e.g., by jamming) and make aircraft rely on backup systems using unauthenticated data links (e.g., CPDLC) where manipulation is much harder to detect.

\protocol*{Controller Pilot Data Link Communications (CPDLC)}

CPDLC is a message-based service offering an alternative for voice communication between ATC and pilot. ATC can use CPDLC via a terminal to send clearances or requests. The pilot can send requests and reports by selecting predefined phrases (e.g., REQUEST, WHEN CAN WE) or free text. CPDLC has great advantages over VHF: the number of acoustic misunderstandings is reduced, messages are saved for accountability, and it is easier, more efficient, and safer to transmit and receive long messages such as flight plan changes during flight. For example, VHF depends on the controller to catch a wrongly understood flight level instruction while he is also busy with several other aircraft; with CPDLC such mistakes can be eliminated and the communication demands on the pilot reduced by as much as 84\% \cite{massimini1999insertion}.

CPDLC is not widely used yet as indicated by the fact that 30\% of our survey respondents had never heard of it. Some busy airports already employ it for automated clearance delivery and start up approval, in some European airspaces it is also used in-flight for minor tasks. 
Currently, CPDLC uses VHF Data Link Version 2 (VDL) \cite{DO-281B} as its data link. Coverage is provided by ground stations and satellites to ensure availability where required, even in oceanic regions. It has been successfully used for more than a decade in airspaces not covered by VHF as it offers easier communication (via satellite) compared to HF with its signal propagation difficulties. Without CPDLC, time-critical ATC clearances or pilot requests very often cannot reach the destination in due time, which forces pilots to deviate from ATC clearance without permission (for example to avoid bad weather) causing a safety problem. Indeed, the transit times of HF Data Link (HFDL) messages do not meet the requirements for some aircraft separation standards. Such issues are eliminated with CPDLC.

\subsubsection*{Security Considerations}
Compared with VHF, relying on an unauthenticated data link becomes a larger problem as attack detection is much harder without voice recognition. Thus, it is problematic that less than 10\% of all respondents are aware that CPDLC is not authenticated, especially considering the plans of many aviation authorities to shift more responsibilities towards CPDLC in the future. As shown in Fig.\,\ref{fig:integrity-impact-atc}, across all groups, a majority of those respondents, who said they knew the answer, was mistaken about its security capabilities.  

Attacks on CPDLC's availability are less critical currently, as CPDLC is still used as a secondary communication layer. Protocol attacks such as message manipulation or injection are severe when undetected, as clearances and other flight safety-related information are transmitted using CPDLC. As there is no authentication, it is trivial to eavesdrop on or spoof clearances and execute replay and message alteration attacks as discussed further in \cite{mahmoud2014aeronautical2}. Impersonation is easily possible: to login to the responsible Area Control Center, the pilot simply puts the correct Location Indicator (e.g., SBAO for ATLANTICO, the control center in Recife, Brasil) into the terminal. After a handshake, the user is successfully logged in. Likewise, an attacker can claim the identity of an ATC unit and send instructions to an aircraft causing the pilot to perform unnecessary, dangerous manoeuvres or causing ATC inquiries.

\protocol*{Primary Surveillance Radar (PSR)}

PSR is the acronym for non-cooperative aircraft localization systems using radar. In aviation, these usually consist of a rotating antenna radiating a pulse position-modulated and highly directional electromagnetic beam on a low GHz band \cite{skolnik2008radar}. The pulses are reflected by targets and bearing and round trip time are measured to get the target's position. PSR is independent of aircraft's equipment, however dependent on the reflecting area (surface material and size, distance and orientation of the aircraft in space). Due to this, and the fact that the signal has to travel two-way, very high radiation power is required (several MW). As the received information is carried by analog signals, the system has to deal with numerous disturbing echoes caused by terrain, obstacles, weather, flocks of birds or even cars on elevated roads. This makes complex signal processing necessary to extract the desired information.
In military airspace surveillance, PSR is strictly required, as it is crucial to detect aircraft with intentionally non-working transponders. In civil ATC, however,\footnote{E.g., in the major Central European airspace one of the authors works in.} PSR is used merely to detect aircraft with rare transponder failures and not as standard backup. Neither identification nor altitude are provided by PSR, the tracker software system uses PSR solely to verify and improve the quality of targets obtained by other sensors.

\subsubsection*{Security Considerations}
As PSR systems use a signal-based detection approach, they are not subject to protocol attacks such as message injection. However, jamming on any of the operational frequencies is possible \cite{center2012electronic}, although due to high power requirements remains in the realm of military electronic warfare. Normally, missing PSR information (caused by jamming) does not impact controllers as the main target information (position, identification, altitude, intent) is provided separately. This might also explain the large majority of controllers (almost 60\%), who believe PSR offers integrity checks. While military PSR can offer security measures such as frequency hopping or modulation schemes, these are unavailable in civil aviation. Similar to other sensor systems, PSR may also be vulnerable to attacks on its timebase (e.g., GPS).  Still, PSR can be considered relatively secure compared to other technologies.
PSR belongs to the oldest ATC technologies and is phased out in favor of more modern data communication protocols using more accurate satellite systems (see below). This can potentially reducing long-term security as reliance on unauthenticated dependent technologies increases.

\begin{table*}[t]
\scriptsize
\begin{tabular}{@{}>{\raggedright}p{2.35cm}>{\raggedright}p{2.92cm}>{\raggedright}p{2.5cm}>{\raggedright}p{2.85cm}>{\raggedright}p{2.75cm}p{2.65cm}l@{}}
\toprule
 & \textbf{Voice} & \textbf{PSR} & \textbf{Mode A/C/S} & \textbf{ADS-B}  & \textbf{CPDLC} \\ \toprule
\textbf{Use} & Communication ATC-Cockpit & Non-cooperative aircraft detection and positioning & Cooperative aircraft detection, positioning and data exchange & Broadcast ATC and collision avoidance relevant aircraft data  & Communication ATC-Cockpit\\ \midrule
\textbf{Type} & Selective \& Broadcast & Broadcast & Interrogation & Broadcast  &  Selective\\ \midrule
\textbf{Sender} & Aircraft \& Ground & Ground & Aircraft & Aircraft & Aircraft \& Ground \\ \midrule
\textbf{Receiver} & Aircraft \& Ground & Original Sender& Aircraft \& Ground & Aircraft \& Ground & Aircraft \& Ground \\ \midrule
\textbf{Frequency} & 3.4-23.35, 117.975-143.975, 225-400\,MHz & 1-2 \& 2-4\,GHz band & 1030 \& 1090\,MHz & 978 \& 1090\,MHz  & VDL2: 136.975\,MHz\\ \midrule
\textbf{Data Rate} & Not applicable & Not applicable & 1\,Mbps (Mode S) & 1\,Mbps & 30\,kbps\\ \midrule
\textbf{Contents} & Clearances, pilot requests, any other information & Pulses & A: squawk, C: altitude, S: as ADS-B but no position & position, vel\-ocity, alt\-itude,  ID, callsign, intent, etc.  & Clearances, requests, weather, further data\\ \midrule
\textbf{Link Layer} & Radio (amplitude modulation) & Pulse position modulation  & Mode A/C/S & UAT / Mode S 1090ES & VDL / HFDL / satcom \\ \midrule
\textbf{Data Source} & Pilot \& Controller & Radar  & Aircraft & Aircraft & Several \\ \midrule
\textbf{Signal} & Analog & Analog & Digital & Digital & VDL2+: digital \\ \midrule
\textbf{Adoption} & In use & In use & In use & Parts of the world, in adoption  &  Parts of the world, in adoption\\ \midrule
\textbf{Standards/References} &\hspace{-0.1bp}\cite{Anx10V3,Anx10V5} &\hspace{-0.1bp}\cite{skolnik2008radar} &\hspace{-0.1bp}\cite{Anx10V4} &\hspace{-0.1bp}\cite{DO-260B,UATsarps, UATspec} &\hspace{-0.1bp}\cite{DO-281B}\\
\bottomrule
\end{tabular}
\caption{Detailed characteristics of air traffic control protocols.}
\label{fig:ATC-table}
\end{table*}



\protocol*{Secondary Surveillance Radar (SSR)}

The transponder modes A, C, and S (short: Mode A/C/S) are part of the Secondary Surveillance Radar \cite{Anx10V4}. The cooperative technology provides more target information on ATC radar screens, as PSR only offers an unidentified target position without further data. SSR uses digital messages with different frequencies and modulations for the interrogation (1030\,MHz) and the reply (1090\,MHz).
SSR ground stations interrogate aircraft transponders, which reply with the desired information. The reply is also used to locate the aircraft's position using the antenna's bearing and the message round trip time. In this digital process, a radiation power of about 1\,kW is sufficient, much less compared to PSR.
The older Modes A and C (ID and altitude) are being substituted by Mode S which supports selective interrogations, to relieve the saturated 1090\,MHz reply channel, suffering from severe message loss. Mode S also offers a worldwide unique transponder ID and further message formats (e.g., aircraft intent or autopilot modes).

\subsubsection*{Security Considerations}
With the publication of Mode S implementations for SDRs on the Internet (e.g., \textit{dump1090}), a somewhat knowledgeable attacker can exercise full control over the communication channel, i.e., by modifying, jamming, or injecting Mode A/C/S messages into ATC systems, he can create a fully distorted picture of the airspace as seen by ATC.

Every Mode S message carries an identifier which can be replaced with an arbitrary one. Using a known and trusted aircraft identifier may, for example, reduce the likelihood for detection  compared to an unexpected object on the radar. Mode S also offers special emergency codes selected by the pilot (e.g., 7500 for hijacking, 7600 for lost communications, and 7700 for an emergency). Injecting these can directly cause ATC inquiries. While this happens occasionally through transponder failures and wrong settings (see \cite{Bloomberg}), this feature can actively be used to create confusion at a busy ground station.

As Mode A/C/S is the only source for all necessary information displayed on ATC radar screen (explaining its high impact rating across all groups), manipulation or jamming is a severe threat since no equivalent backup exists. Mode S messages and its data link are also used by other ATC systems, which consequently inherit most of its vulnerabilities as we will discuss in the next sections. Worryingly, more than 40\% of all respondents wrongly believe the protocol offers built-in security features, including a notable 60\% among controllers.

A vulnerability specific to Mode S is the amplification attack. Exploiting interrogations in Mode S, an attacker can cause large-scale interference on the 1090\,MHz channel without sending on the target frequency (but on the 1030\,MHz interrogation frequency instead). Interrogations are limited to a maximum of 250/s now \cite{allcalls}, but these restrictions are placed on the interrogators, not on the Mode S transponders in aircraft. As Mode S messages are unauthenticated, a malicious sender can easily circumvent these measures and use non-selective interrogations to amplify her sending power and frequency. By changing her own identifier code, the attacker can make all receiving aircraft answer the interrogations, increasing the range and capability of the interference attack manifold. While the aircraft continue to send useful information, the interference level in even moderately busy airspaces would quickly cause a partial DoS as important data gets lost.

As a concrete example, the FAA uses a stochastic probability of 0.25\footnote{Accounting for interference and the time when a transponder is busy.} to calculate the number of replies to a given Mode S all-call \cite{allcalls}. Thus, an attacker can create an amplified response of $x*y*0.25$  messages per second, where $x$ is the number of receiving aircraft and $y$ the number of all-calls per second. Based on data from our own Mode S receivers, 200 aircraft are easily in range within a normal airspace. Thus, an attacker can use only 100 messages to create $100*200*0.25=5000$ additional Mode S messages, adding to the significant existing interference experienced by all ATC receivers.\footnote{The signal load is approaching 100\% in some scenarios \cite{acasa}.} If done with consideration, such an attacker is also more difficult to detect from the ground. In addition, a recent incident showed that some transponders are susceptible to over-interrogation. Sending many interrogation calls can lead to overheating and result in a complete loss of the target from ATC displays due to a full DoS of the transponders. Investigations revealed that transponders transmit at rates beyond their requirements and design limits, worsening amplification attacks \cite{EASA15}.

\protocol*{Automatic Dependent Surveillance - Broadcast (ADS-B)}
The ADS-B protocol \cite{DO-260B} is used by aircraft to continually broadcast their own ID, position and velocity as well as further information such as intent or urgency codes. These broadcasts happen twice a second in case of position and velocity, and once every 5\,s for identification. It is mandated for use by 2020 in US and European airspaces, promising to improve on location accuracy and decrease system costs by replacing SSR  \cite{strohmeierIEEEcomsurv}. ADS-B exemplifies the move to cooperative data communication networks in the next ATC protocol generation but as of now, its impact is limited compared to VHF and SSR.

\begin{figure*}[t]
\centering
\includegraphics[bb=14bp 2bp 390bp 285bp,clip,width=0.446\textwidth,height=130bp]{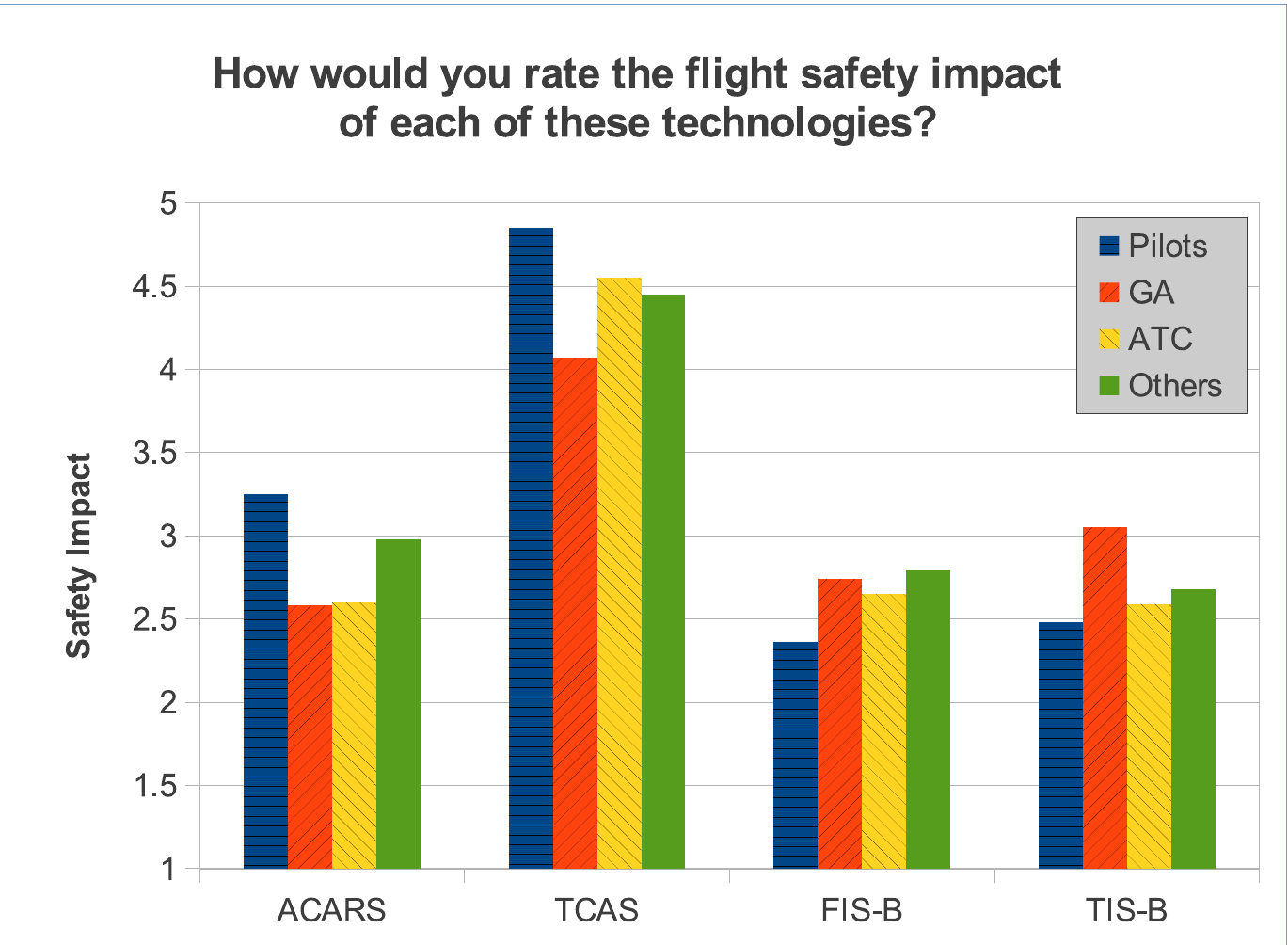}\includegraphics[bb=23bp 10bp 555bp 360bp,clip,width=0.476\textwidth,height=135bp]{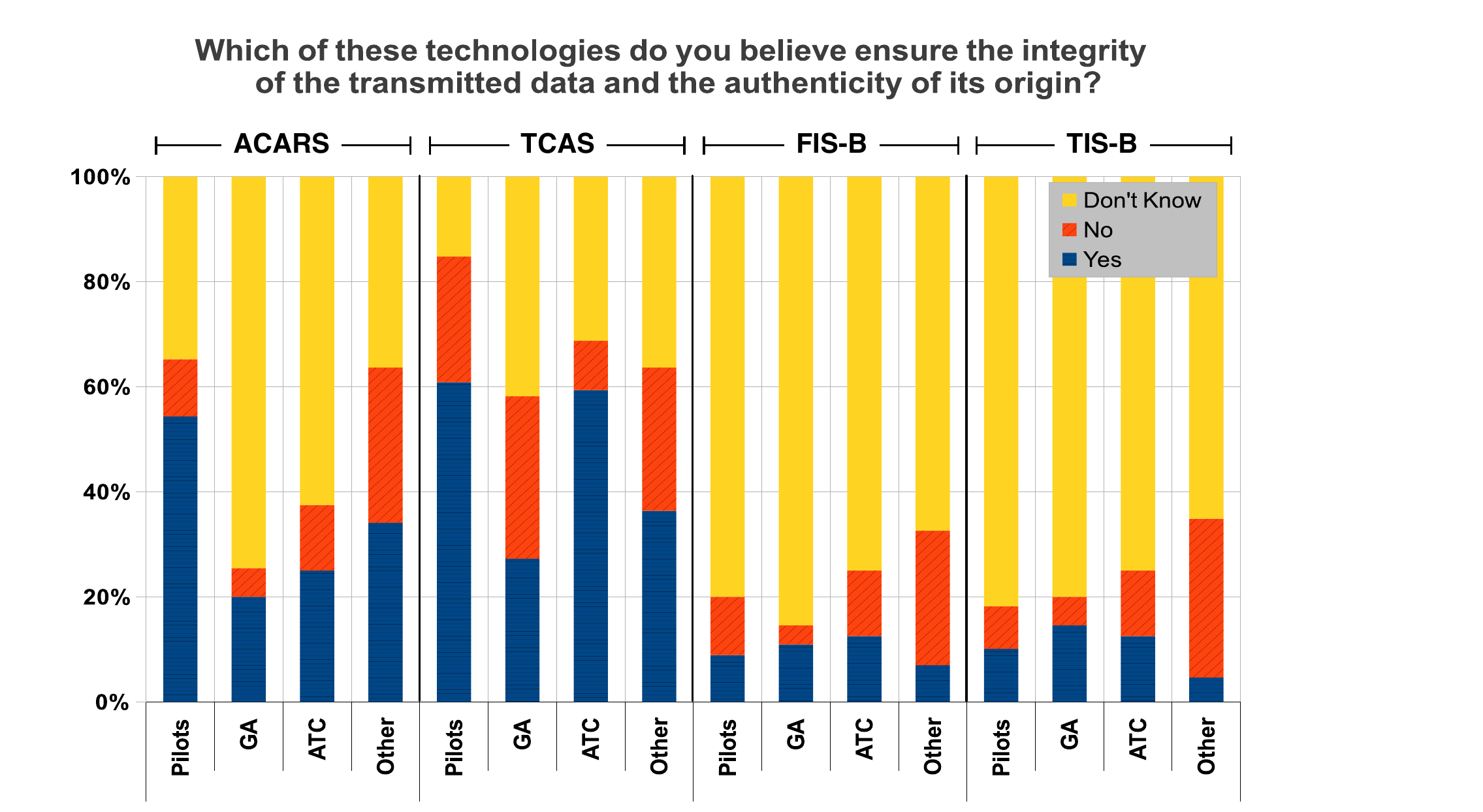}
\caption{Assessment of information services' impact and security capabilities (authentication and integrity).}
\label{fig:integrity-impact-inf}
\vspace{-5pt}
\end{figure*}

\subsubsection*{Security Considerations}

As the commercially used ADS-B data link 1090 Extended Squitter (1090ES) is based on unauthenticated Mode S, it suffers from the same passive and active attacks. For example, it is possible to selectively jam all ADS-B messages of a single aircraft, which would make it vanish from the ADS-B channel. This feat is much more easily accomplished with ADS-B's regular broadcasts compared to Mode S' interrogation system using directional antennas and much more frequent, irregular interrogations.

Furthermore, as ADS-B additionally broadcasts the position of aircraft, this opens some new attack vectors (we do not go into detail as ADS-B has been the focus of recent research and media attention, see, e.g., \cite{McCallie2011,Purton2010}) which only require standard off-the-shelf hardware to execute as demonstrated in \cite{Costin,Schaefer13}. Trivially injected ADS-B messages claiming to be non-existing aircraft are impossible to tell apart from authentic ones on the link layer. Other attacks virtually modify the trajectory of an aircraft by selectively jamming an aircraft's messages and replacing them with modified data. This causes discrepancies between the real position and the one received by ATC \cite{Schaefer13}. This is a worrying prospect, as ADS-B is set to be the main ATC protocol in the long term, with the FAA considering elimination of Mode A/C/S transponders at some point in the future.\footnote{See https://www.faa.gov/nextgen/programs/adsb/faq/} Yet, less than 20\% of the participating pilots and controllers are aware of its shortcomings.

\protocol*{Multilateration (MLAT)}
Multilateration, or hyperbolic positioning, has been successfully employed for decades in military and civil applications, not limited to navigation. It differs from other ATC aids as it is not a separate protocol but exploits the time differences of arrival of signals received from aircraft independently using other protocols, SSR or ADS-B in the aviation context \cite{Wood98multilateration}. Using the reception times of four or more receivers, it is a purely geometric task to find the unknown point. Fig.\,\ref{fig:integrity-impact-atc} shows that there is a significant difference in the impact assessment between pilots, who do not actively use MLAT, and controllers whose work partly relies upon it. This also explains the high uncertainty surrounding the security features of MLAT in the commercial/private pilot groups.

\subsubsection*{Security Considerations}
In theory, MLAT as a technology does not rely on the \textit{contents} of the received messages but (similar to PSR) works purely on the signal level. This provides a strong theoretical security advantage because it does not suffer from compromised message integrity. Even if the contents of, e.g., an ADS-B message are wrong, the location of the sender can still be identified. Thus, MLAT offers additional security based on physical layer properties (here the propagation speed of electromagnetic waves) which are difficult to cheat.

However, in practical implementations this assumption fails. Typical MLAT systems heavily rely on fusing the location data gained from the signals with SSR message contents to display identification and altitude of the targets, leaving the system as a whole vulnerable. Independently from this problem, a well-coordinated and synchronized attacker could still manipulate a message's time of arrival at the distributed receivers of an MLAT system such that using these signals for location estimation would result in a position of the attacker's choice \cite{MoserThesis}. This is shown in \cite{Tippenhauer11} for the similar case of spoofing a group of distributed GPS receivers. The authors find that even though more receivers severely restrict the possible attacker placement, attacks are generally feasible. Yet, MLAT is a favored SSR backup solution in aviation \cite{icaoadsbsecurity} and academic communities \cite{Sampigethaya2011}. Unfortunately, it is very expensive to deploy (in part due to its susceptibility to multipath propagation) and thus not a preferred option in all environments \cite{strohmeier2015lightweight}.

\begin{table*}[t]
\scriptsize
\begin{tabular}{@{}>{\raggedright}p{2.4cm}>{\raggedright}p{4cm}>{\raggedright}p{3.5cm}>{\raggedright}p{3.5cm}>{\raggedright}p{2.8cm}l@{}}
\toprule
 & \textbf{ACARS} & \textbf{TCAS} & \textbf{FIS-B} & \textbf{TIS-B} & \textbf{} \\ \toprule
\textbf{Use} & Dispatch, operations, maintenance... & Collision avoidance & Flight information & Traffic information &  \\ \midrule
\textbf{Type} & Broadcast & Interrogation & Broadcast & Broadcast &  \\ \midrule
\textbf{Sender} & Aircraft \& Ground & Aircraft & Ground Radar & Ground Radar &  \\ \midrule
\textbf{Receiver} & Aircraft \& Ground & Aircraft \& Ground & Aircraft & Aircraft &  \\ \midrule
\textbf{Frequency} & 129.125-136.900\,MHz & 1030 \& 1090\,MHz & 978\,MHz & 978 \& 1090~MHz &  \\ \midrule
\textbf{Data Rate} & 2400~bps & 1~Mbps & 1~Mbps & 1~Mbps &  \\ \midrule
\textbf{Contents} & Position, weather, fuel \& engine information, delays, maintenance... & Altitude, relative position, transponder status & Weather text \& graphics, notices to airmen, terminal information & Non-ADS-B equipped aircraft &  \\ \midrule
\textbf{Link Layer} & Several & Mode S \& 1090ES & UAT & UAT \& 1090ES &  \\ \midrule
\textbf{Data Source} & Various & ADS-B \& Mode S & FIS-B Provider & Radar station &  \\ \midrule
\textbf{Signal} & Digital & Digital & Digital & Digital &  \\ \midrule
\textbf{Adoption} & In use & In use & Parts of the US & Parts of the US &  \\
\midrule
\textbf{Standards/References} &\hspace{-0.1bp}\cite{ARINC618-7} &\hspace{-0.1bp}\cite{DO-185B} &\hspace{-0.1bp}\cite{DO-358} &\hspace{-0.1bp}\cite{DO-260B} & \\ 
\bottomrule
\end{tabular}
\caption{Detailed characteristics of information services protocols.}
\label{fig:INF-table}
\vspace{-15pt}
\end{table*}

\subsection{Information Services}
Information services are more general platforms for the exchange of aviation information. They use a variety of sources and supply the backbone for a wide array of use cases. Table \ref{fig:INF-table} provides the technical details, and Fig.\,\ref{fig:integrity-impact-inf} gives the survey respondents' assessment of their safety impact and security features. Overall, we can see that TCAS is considered to have a very high safety impact by all respondents, while the other information services are less important, at least towards safety.


\protocol*{Aircraft Communications Addressing and Reporting System (ACARS)}
ACARS \cite{ARINC618-7} is a digital datalink developed for general communication between aircraft and ground stations. ACARS messages are divided into two types, air traffic services messages for ATC, flight information and alerting, and aeronautical operational control messages which most airlines use to communicate with their aircraft. It is used in all flight phases, for services as varied as dispatch, operations, engineering, catering, or customer service. They transmit safety-critical data such as aircraft weight, fuel, engine data, or weather reports; privacy-related information about passengers or catering requests; and information critical to business operations such as gate assignments, crew schedules, or flight plan updates.

ACARS offers five data links, depending on the aircraft's equipment: VHF, Inmarsat satcom, Iridium satellite, VDL Version 2, and High Frequency Data Link. The messages are character-oriented and only accept valid ASCII symbols \cite{spitzer2014digital}.

\subsubsection*{Security Considerations}
ACARS security issues are long known, a 2001 military study says \cite{risley2001experimental}: ``The military is uncomfortable with the ease at which eavesdropping on ACARS can be achieved.'' Today, ACARS eavesdropping has become much more widely accessible as SDR-based decoders are available on the Internet (e.g., \url{http://acarsd.org}), and even the satellite data link is easy to attack. To counter this, the ACARS message security (AMS) standard was developed \cite{ARINC823-1}. It provides end-to-end encryption using ECDSA with SHA256 for digital signatures and offers message authentication codes with HMAC-SHA256 of a default length of 32 bits.
AMS currently enjoys very little adoption as only few airlines (e.g., Lufthansa \cite{lufthansaacars}) even consider securing ACARS transmissions. Others (e.g., Ryanair \cite{runwaygirl}) forego ACARS completely and use airport-based mobile phone technologies.

Furthermore, airlines use their own semantics for data packets transmitted by ACARS, providing some security by obscurity.\footnote{A US military presentation \cite{risley2001experimental} considers binary AOC messages less vulnerable as they are not human-readable.} Due to these differences in implementations and applications supported by ACARS, discussing all potential attack vectors is not possible here. Yet, there are some examples in the literature, including the potential exploitation of soft- and hardware flaws using the interface offered by ACARS \cite{teso2013aircraft} or issuing wrong ATC instructions \cite{risley2001experimental}. Besides, it is easy to imagine the impact on business intelligence and personal privacy when passenger lists, crew information, or engine data are transmitted in clear text via ACARS.

\protocol*{Traffic Alert and Collision Avoidance System (TCAS)}
TCAS \cite{DO-185B} is an airborne system for collision avoidance independent of ground-based ATC. The current version TCAS II uses the available information (i.e., identity, altitude) from ATC protocols such as Mode C and S (ADS-B will be incorporated in the future) to provide a traffic surveillance display of all equipped aircraft in the proximity \cite{spitzer2014digital}. It determines the relative velocity and distance of nearby transponder-equipped aircraft through interrogation. When a broadcast Mode S message is received, the transmitted ID is added to a list of aircraft that is then interrogated at about 1\,Hz. With the reply, distance and altitude of the interrogated aircraft are determined. With ADS-B messages, the interrogation step becomes unnecessary. Based on the relative velocities and positions, potential threats are identified and presented to the  pilot as a Traffic Advisory. When proximity thresholds are violated, TCAS issues a Resolution Advisory (RA) and proposes an avoidance maneuver to eliminate the threat (in the latter case TCAS can be classified as ATC protocol, too). Due to its function, TCAS has received one of the highest safety impact ratings from all our respondents, particularly the commercial pilots (see Fig.\,\ref{fig:integrity-impact-inf}).

\subsubsection*{Security Considerations}
As TCAS is based on data and message formats of Mode A/C/S (and ADS-B, which it will integrate in newer versions), it suffers from the unauthenticated nature of these protocols described above. Strikingly, only 25\% of our respondents realized this as shown in Fig.\,\ref{fig:integrity-impact-inf}, including less than 10\% of the ATC group. The potential attack vectors of TCAS differ, however, as the main targets are aircraft, not ground stations. Attacking aircraft at cruising altitude requires a strong transceiver, making an attack from within the aircraft more likely.
One concern is an attacker who falsifies the data that TCAS uses to be aware of the surveillance picture around an aircraft. To do this, answers to Mode S interrogations by TCAS are spoofed using wrong information and message timings. The attacked TCAS system will classify such ghost aircraft as a threat and initiate an RA to which the pilot needs to respond. As even real advisories can lead to serious incidents, a loss of situational awareness is very possible \cite{easa-safety}.


Another attack focuses on the RA messages themselves. Issuing fake advisories to ATC ground stations using Mode S RA reports is an easy way to cause a partial loss of situational awareness and control. Since controllers are prohibited from interfering with RAs, effective air traffic controlling is strongly inhibited. Considering the relative ease of such attacks, and the potentially severe impact, TCAS provides the biggest contrast of all protocols between the participants' security perception and the technical reality, only matched by Mode A/C/S.

\begin{figure*}[t]
\centering
\includegraphics[bb=14bp 2bp 390bp 285bp,clip,width=0.446\textwidth,height=140bp]{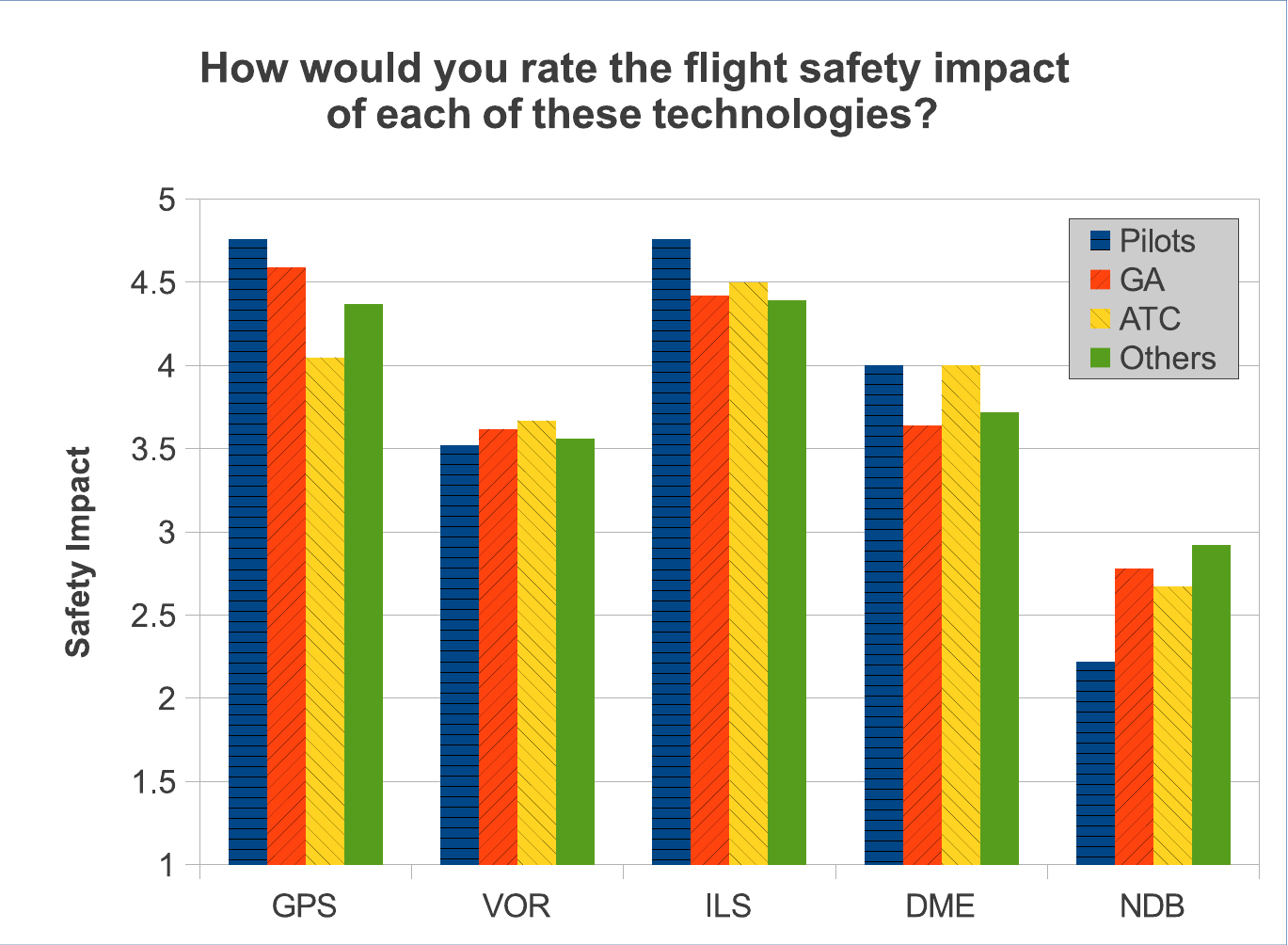}\includegraphics[bb=23bp 10bp 535bp 360bp,clip,width=0.476\textwidth,height=145bp]{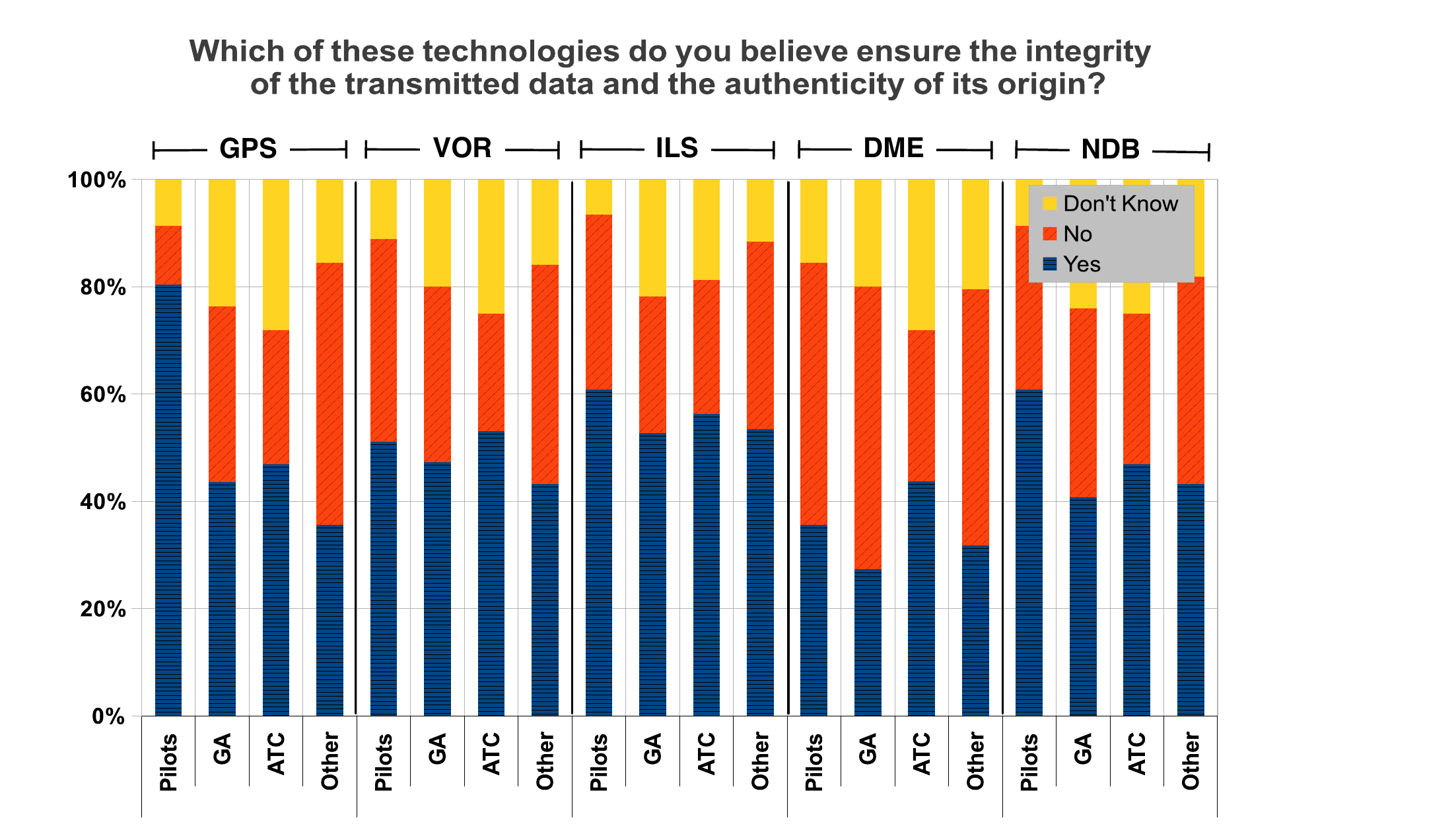}
\caption{Assessment of navigation aids safety impact and security capabilities (authentication and integrity).}
\label{fig:integrity-impact-nav}
\end{figure*}

\protocol*{Flight Information System - Broadcast (FIS-B)}
FIS-B \cite{DO-358} is a general flight information service that requires aircraft to be equipped with ADS-B In. It uses the Universal Access Transceiver (UAT) \cite{UATsarps, UATspec} datalink on 978\,MHz which offers more flexibility through larger ADS-B messages. It is in use in parts of the US, with a wider adoption possible in the future (many  European respondents did not know the protocol). FIS-B provides data about airspace restrictions, or meteorological advisories. The data is supplied by the FAA for general aviation below 24,000~ft \cite{sbsdescription}.

\subsubsection*{Security Considerations}
FIS-B is based on the unauthenticated ADS-B data link UAT. It is thus trivial to manipulate or replay the broadcast messages sent out by ground station service providers \cite{DO-267A}.  The payload encoding of FIS-B is available at http://fpr.tc.faa.gov, requiring a non-verified registration. Decoders of weather data sent over FIS-B such as METAR (Meterological Aviation Reports) are widely available on the Internet. 
Thus, it is not difficult to send out forged broadcasts of weather reports or severe weather forecast alerts, even raster scan pictures, following the standard specification.

\protocol*{Traffic Information System - Broadcast (TIS-B)}
TIS-B \cite{DO-260B} is another ground-based traffic information service used in the US that broadcasts additional data about aircraft that are not equipped with ADS-B transponders. TIS-B is used for increased situational awareness and collision avoidance. The system uses the same frequencies as ADS-B and the same message format and provides users with a full surveillance picture as seen by ground radar, i.e., the broadcast data can be compiled from all available sources such as PSR, SSR, ADS-B, or MLAT.

\subsubsection*{Security Considerations}
TIS-B uses both available ADS-B data links. Again, as both are unauthenticated, it is trivial to manipulate or replay the broadcast messages sent out by ground station service providers with the same means as explained above. Forged TIS-B messages broadcast to airborne targets can advertise non-existing aircraft or manipulate information (e.g., position) about  aircraft without their own beacon transponder \cite{sbsdescription}. Both FIS-B and TIS-B are currently of limited safety impact outside areas busy with general aviation in the US, which explains why many of our participants do not have an informed opinion about their security features.

\subsection{Radio Navigation Aids}\label{sub:NavigationAids}

This section provides a brief introduction to the most common radio navigation aids. More detailed descriptions can be found in the ICAO standard document \cite{Anx10V1} or in \cite{Kayton97}. An overview of the technical details is provided by Table \ref{fig:NAVAID-table}. Due to their much simpler application - to help navigate - they are also subject to simpler attacks, which we discuss together.

\protocol*{Global Positioning System (GPS)}
GPS \cite{GPSspec} is the de facto standard global navigation satellite system (GNSS). It allows for determining the position and velocity of its users and precise time coordination. It is a ranging system from known positions of multiple satellites to an unknown position on land, sea, and air. Orbiting satellites therefore continuously transmit navigation signals which include precise timestamps and information to reduce the positioning error. The ranging is done by measuring the time of flight or the relative phase offsets of these signals from $n\ge4$ satellites. This results in a system of $n$ equations with four unknowns (the three-dimensional coordinates and the offset of the user's clock to the system time) which can be solved using numerical methods to obtain the desired position and time information \cite{Kayton97}.
The use of GPS for air navigation is only allowed if the integrity of the derived information is monitored at all times \cite{2015faraim}. There are two basic approaches: In ground- or satellite-based augmentation systems (GBAS or SBAS, respectively), fixed ground stations monitor the integrity of the GPS signals and notify users about errors. This integrity information is then propagated to the users directly (i.e., ground-based) or via satellites.

\begin{table*}[t]
\scriptsize
\begin{tabular}{@{}>{\raggedright}p{2.3cm}>{\raggedright}p{2.8cm}>{\raggedright}p{2.8cm}>{\raggedright}p{2.75cm}>{\raggedright}p{2.75cm}p{2.85cm}@{}}
\toprule

 & \textbf{GPS} & \textbf{VOR} & \textbf{ILS} & \textbf{NDB} & \textbf{DME} \\ \toprule
\textbf{Use} & Positioning, Time & Bearing & Approach Guidance & Direction & Distance \\ \midrule
\textbf{Type} & Broadcast & Broadcast & Broadcast & Broadcast & Interrogation \\ \midrule
\textbf{Sender} & Satellite & Ground Station & Ground Antenna Array & Ground Station & Ground Station \\ \midrule
\textbf{Frequency} & 1.57542 \& 1.2276\,GHz & 108.975-117.975\,MHz & 75, 108-111.975, 328.6-335.4\,MHz & 190-1,750\,kHz & 962–1,213\,MHz \\ \midrule
\textbf{Contents} & 3D position, time sync & Bearing & Deviation from approach & Direction to ground station & Distance to ground station \\ \midrule
\textbf{Method} & Time difference of arrival & Phase shift & Signal strength & Angle of arrival & Round-trip time \\ \midrule
\textbf{Standards/References} &\hspace{-0.1bp}\cite{GPSspec,vieweg1993aircraft} &\hspace{-0.1bp}\cite{Anx10V1,Kayton97} &\hspace{-0.1bp}\cite{Anx10V1,Kayton97} &\hspace{-0.1bp}\cite{Anx10V1,Kayton97} & \hspace{-0.1bp}\cite{Anx10V1,Kayton97} \\  \bottomrule
\end{tabular}
\caption{Detailed characteristics of standard radio navigational aids.}
\label{fig:NAVAID-table}
\end{table*}

An alternative approach to integrity monitoring is to use an aircraft-based augmentation system. Here, the position solution is integrated with redundant information available on board the aircraft. There are two general classes of aircraft-based integrity monitoring: receiver autonomous integrity monitoring (RAIM), which uses GNSS information exclusively, and aircraft autonomous integrity monitoring, which uses information from additional on-board sensors (e.g., barometric altimeter, clock or the inertial navigation system) \cite{vieweg1993aircraft}. In RAIM, at least six satellites are used to detect faulty satellite signals by solving the system of equations for all subsets of the satellites.

\protocol*{VHF Omnidirectional Radio Range (VOR)}
VOR \cite{Anx10V1} enables aircraft to determine their clockwise angular deviation (bearing) from magnetic North to fixed ground stations. The ground stations broadcast an omnidirectional reference signal and a highly directional variable signal. The reference signal is a frequency modulated 30~Hz signal with constant amplitude. The variable signal is a signal which rotates clockwise in space with 30 rotations per second. This rotation results in a 30~Hz amplitude modulation component of the carrier frequency. Both signals are synchronized such that their phase-offset corresponds to the direction of transmission. This offset can be determined at a receiver by comparing the phase of the frequency-modulated reference wave with that of the amplitude-modulated signal component.

To identify a VOR station, its two- or three-letter identifier is broadcast using Morse code. In most cases, the identifier along with advisories and service information are also transmitted in a recorded automatic voice.\footnote{This depends on the equipment of sender/receiver, see \cite{2015faraim}, Section 3.3.6.} To use VOR for navigation, the stations' positions must be known. If two or more VORs are received, the intersection of the lines of positions to the VORs is the receiver's position.


\protocol*{Nondirectional Beacons (NDB)}
NDBs \cite{Anx10V1} are omnidirectional transmitters at fixed positions. Aircraft equipped with direction-finders (rotating directional antenna or beamforming-like antenna arrays) can determine the angle of arrival of the signal of an NDB and thereby the direction of the transmitter. A Morse code identifier is sent every 30\,s using on/off keying. NDBs can be used for navigation by looking up their position in a chart.

\protocol*{Distance-measuring Equipment (DME)}
DME \cite{Anx10V1} is an interrogation-based ranging system. Aircraft transmit a random sequence of pulse pairs and the ground transponder responds after a fixed delay (typically 50\,µs) with the same pulse sequence. The interrogating aircraft searches for this pattern and locks to it to have a narrowed search window for the next interrogations. The distance from the aircraft to the transponder corresponds to the round-trip time (minus the 50\,µs) multiplied with the propagation speed of the signals.

\protocol*{Instrument Landing System (ILS)}
ILS \cite{Anx10V1} consists of several (typically four) radio transmitters to guide aircraft during approach and landing.  The localizer antenna is centered on the runway beyond its end. Using amplitude modulation, a 90\,Hz and a 150\,Hz tone are sent with directional antennas in two lobes. One lobe is slightly directed to the left hand of the centerline and one to the right hand. Thus, a receiver not located on or very close to the front course line will hear one of the signals louder than the other. The localizer provides lateral guidance to the aircraft which keeps flying where both signals are equally loud. Vertical guidance is provided by the glide scope antenna beside the runway using the same principle. Along the approach path, marker beacons may radiate fan-shaped vertical beams at fixed distances to the runway. The beams modulate Morse code-like audible tones with different frequencies, which are provided to the pilot acoustically or visually. The frequency and the Morse code indicate the distance to the runway. The use of marker beacons, however, is rare since real estate for installation is a major problem \cite{Kayton97}; DME is usually used as a replacement.

\subsubsection*{Security Considerations for Radio Navigation Aids}
Fig.\,\ref{fig:integrity-impact-nav} provides the comparative safety impact of all 5 technologies. The differences reflect their current use and importance: GPS and ILS obtain high ratings especially from pilots, DME and VOR follow with middling ratings across all groups, while the outdated NDB is considered to be of low impact. 

Navigational aids measure physical properties of signals such as time-difference of arrival (GPS), angle of arrival (NDB), phase shifts (VOR), signal strength (ILS), or round-trip time (DME). Neither property nor the signals themselves are authenticated and can therefore be trivially attacked within the line of sight. While this has long been achievable by military actors (see existing literature on electronic warfare, e.g., \cite{adamy2001ew}), today many attacks on navigational aids can be launched by less powerful attackers using SDRs. For instance, all radio frequency-based technologies are inherently susceptible to jamming attacks. Hence, a DoS attack only requires the attacker to be able to generate noise such that the signal-to-noise ratio of legitimate signals at the aircraft drops below the reception threshold. GPS signals in particular are extremely weak and easy to attack, a well-known fact in the security community for many years \cite{warner2002simple}. In fact, even some participants experienced failure of their GPS systems due to \emph{``large-scale GPS jamming, presumably local military exercise''}. 

Spoofing is a more surgical but serious threat to navigational aids. As there is only one victim in most attack scenarios (the target aircraft), it is feasible using a single antenna (as shown in \cite{Tippenhauer11} for the case of GPS) and portable implementations of GPS spoofing devices exist \cite{humphreys2008assessing}. RAIM does not provide protection, since the attacker can spoof all six satellites -- it is a safety precaution, not a security feature. Such spoofing attacks are dangerous as they result in a derivation of a false position by the aircraft, which is much harder to detect than a DoS. Existing simulators such as the Spirent GSS7700 make it convenient to generate signals in advance which can be used to spoof a whole track, leading to potentially serious incidents during GPS-based approaches.

\begin{table*}[t]
\begin{tabular}{|>{\raggedright}m{6.2cm}||>{\centering}m{1.05cm}||>{\centering}m{1.25cm}|>{\centering}m{1.82cm}|>{\centering}m{1.83cm}|>{\centering}m{1.8cm}|>{\centering}m{1.3cm}|}
\hline 

{\small {}\textbf{What is the effect if...}} & {\small {}Surveyed group} & {\small {}No effect} & {\small {}Minor loss of situational awareness} & {\small {}Major loss of situational awareness} & {\small {}Full denial of aviation service} & {\small {}Don't know}\tabularnewline
\hline 
\hline 
{\small {}1)\,...a non-existing target shows up on an air traffic control radar
screen? } & {\small{}ATC} & {\small{}16.13\%} & {\small{}54.84\%} & {\small{}25.81\%} & {\small{}0..00\%} & {\small{}3.23\%}\tabularnewline
\hline 
{\small {}2)\,...a non-existing target shows up on a TCAS screen? } & {\small{}Pilots} & {\small{}10.75\%} & {\small{}44.09\%} & {\small{}31.18\%} & {\small{}0.00\%} & {\small{}13.98\%}\tabularnewline
\hline 
{\small {}3)\,...wrong label indications show up on an air traffic control
radar screen (e.g.,\,altitude, selected altitude, callsign)? } & {\small{}ATC} & {\small{}3.01\%} & {\small{}28.31\%} & {\small{}54.82\%} & {\small{}6.02\%} & {\small{}7.83\%}\tabularnewline
\hline 
{\small {}4)\,...wrong label indications show up on a TCAS screen (e.g. relative
altitude)? } & {\small{}Pilots} & {\small{}3.23\%} & {\small{}30.11\%} & {\small{}51.61\%} & {\small{}4.30\%} & {\small{}10.75\%}\tabularnewline
\hline 
{\small {}5)\,...information or whole targets are selectively missing from
an air traffic control radar screen? } & {\small{}ATC} & {\small{}6.67\%} & {\small{}6.67\%} & {\small{}83.33\%} & {\small{}3.33\%} & {\small{}0.00\%}\tabularnewline
\hline 
{\small {}6)\,...information or whole targets are selectively missing from
a TCAS screen? } & {\small{}Pilots} & {\small{}6.52\%} & {\small{}23.91\%} & {\small{}48.91\%} & {\small{}7.61\%} & {\small{}13.04\%}\tabularnewline
\hline 
{\small {}7)\,...all data is missing from an air traffic control radar screen
system (for example for 10 minutes)? } & {\small{}ATC} & {\small{}3.23\%} & {\small{}6.45\%} & {\small{}45.16\%} & {\small{}45.16\%} & {\small{}0.00\%}\tabularnewline
\hline 
{\small {}8)\,...all data is missing from a TCAS screen system (for example for 10 minutes)? } & {\small{}Pilots}  & {\small{}6.52\%} & {\small{}27.17\%} & {\small{}45.65\%} & {\small{}6.52\%} & {\small{}14.13\%}\tabularnewline
\hline 
{\small {}9)\,...the position of an aircraft as shown to its pilot is incorrect? } & {\small{}Pilots} & {\small{}5.38\%} & {\small{}22.58\%} & {\small{}60.22\%} & {\small{}7.53\%} & {\small{}4.30\%}\tabularnewline
\hline 
\end{tabular}\protect\caption{Answer distribution for nine hypothetical scenarios as answered by 92 pilots and 31 controllers, respectively. } \label{ImpactAssessmentResults}
\end{table*}

The threat of attacks on navigational aids exists during all phases of flight. During take-off, approach, landing, and terminal, ground-based attackers benefit from low altitudes of aircraft in terms of a lower required transmission power and line of sight to the victim. However, aircraft in the en-route airspace are also vulnerable by transmitters on board. Interestingly, only around 30\% of the survey participants know about lacking data integrity and authenticity checks in navigation aids, with the exception of NDB which is distrusted by about 50\%. For the other four technologies, about 50\% believe that there is built-in security, a worrying number when contrasted with the high impact ratings of GPS and ILS.

When considering the security of navigational aids, the procedural use of these aids must also be taken into account. In most airspaces in the world, pilots are required to rely on at least two means of navigation. Pilots usually have even four or more navigational aids available which they use as backup and for integrity checking. A typical example described by one of the respondents is using the aids with degrading accuracy, i.e., GPS is backed up by VOR/DME, backed up by visual navigation (in clear weather), then dead reckoning (i.e., using the inertial system), and lastly NDB. ILS is usually cross-checked by the pilot with the vertical rate, if the rate does not match the designated three degree slope, the pilot abandons the approach. Due to this procedure, an attacker would have to attack multiple systems simultaneously to avoid being detected. However, attacking single navigational aids may still go undetected and/or result in a more complex situation for the pilot increasing the risk for human failure.

\section{Assessment of Concrete Scenarios}\label{sub:Impact}

To further improve the understanding of insecure wireless aviation technologies, we transformed some of the previously discussed hypothetical attacks into concrete scenarios and evaluated the respondents' assessments of the practical impact of realistic threats. In nine hypothetical scenarios the respondents had to rate the impact of air traffic communication technologies misreporting the current air traffic picture. Note, that this could happen for different reasons, many of which are not caused by an attack. We were interested in the impact regardless of the underlying cause which would typically be abstracted away for users of these systems. The respondents assessed these scenarios with five options: ``No effect'', ``Minor loss of situational awareness'',\footnote{Situational awareness is a concept widely used in aviation despite some criticism. A historically accepted definition is ``the perception of elements in the environment within a volume of time and space, the comprehension of their meaning, and the projection of their status in the near future'' \cite{endsley1995toward}.} ``Major loss of situational awareness'', ``Full denial of aviation service'' and ``Don't know''. Table \ref{ImpactAssessmentResults} shows the questions and the distribution of the results.

Some participants would have liked more information as answers can depend on many specifics. Unfortunately, it is impossible to cover the large number of potential combinations of systems and situations separately. Thus, we consider the quantitative part of the questions about these scenarios as an abstract expert opinion. Consequently, we complemented all questions with a qualitative approach by inviting comments, some of which we provide to gain a more complete picture. We focus on three systems: ATC radar, showing the air traffic picture to ground controllers, (Questions 1,3,5,7), TCAS screens displaying intruders in an aircraft's immediate airspace to the pilot (Q2,4,6,8), and navigation aids (Q9). In terms of attack classes, we analyze message injection (Q1-2), content manipulation (Q3-4,9), selective jamming (Q5-6), and full DoS caused by jamming (Q7-8). We report the answers of 92 pilots to the scenarios relating to their expertise (Q2,4,6,8,9), and of 31 controllers to the scenarios relevant to ATC (Q1,3,5,7).

\begin{table*}[t]
\scriptsize
\begin{tabular}{@{}>{\raggedright}p{2.5cm}>{\raggedright}p{2.9cm}>{\centering}P{1.7cm}>{\centering}P{1.7cm}>{\raggedright}p{3.12cm}l@{}}
\toprule

 & \textbf{Attacks} & \multicolumn{2}{c}{\textbf{Requirements}} &  \multicolumn{2}{c}{\textbf{Defenses}} \\ \hline
 
 & &  \textbf{Confidentiality} & \textbf{Integrity} & \textbf{Long-term} & \textbf{Short-term} \\ \toprule
 
\textbf{VHF} & \hspace{-0.1bp}\cite{stelkens2015towards}  & - & X & \textbf{-}  &  \hspace{-0.1bp}\cite{stelkens2015towards,prinz2005s,hering2003safety,hagmuller2004speech,neffe2007speaker,fantacci2009secure} \\ \midrule

\textbf{PSR} & \hspace{-0.1bp}\cite{center2012electronic,adamy2001ew,adamy2004ew,adamy2008ew} & - & X & \textbf{-}  &  \hspace{-0.1bp}\cite{adamy2001ew,adamy2004ew,adamy2008ew} \\ \midrule

\textbf{SSR} & \hspace{-0.1bp}\cite{EASA15,middleton2012risk} & - & X & \hspace{-0.1bp}\cite{jochum2001encrypted,kenney2008secure,olive2001efficient} & \hspace{-0.1bp}\cite{middleton2012risk,strohmeier2015intrusion} \\ \midrule

\textbf{ADS-B / TIS-B / FIS-B} & \hspace{-0.1bp}\cite{Costin,McCallie2011,Schaefer13,teso2013aircraft,defcontruth,magazu2012exploiting} & - & X &  \hspace{-0.1bp}\cite{wesson2014can,strohmeierIEEEcomsurv,lee2015ads,kacem2015integrity,kacem2015key,kenney2008secure,yang2014ebaa,jochum2001encrypted,yeste2015ads} & \hspace{-0.1bp}\cite{strohmeier2015lightweight,icaoadsbsecurity,Nuseibeh2009,strohmeierIEEEcomsurv, McCallie2011,Purton2010,Smith06,wu2015jamming,wang2015low,yangnew,hableel2015protect,monteiro2015detecting2,monteiro2015detecting,ghose2015verifying,schafer16,schafer2015secure,murphy2014device,leonardi2014ads,strohmeier2015intrusion,park2014high} 
\\ \midrule

\textbf{CPDLC} & \hspace{-0.1bp}\cite{mahmoud2014aeronautical} & X & X &  \hspace{-0.1bp}\cite{mcparland2001securing,patel2015icao,storck2013benefits,mahmoud2014aeronautical,getachew2005elliptic,olive2001efficient} & \textbf{-} \\ \midrule

\textbf{MLAT} & \hspace{-0.1bp}\cite{MoserThesis} & - & X &  \hspace{-0.1bp}\cite{Song2008,Chiang2012} & \hspace{-0.1bp}\cite{Capkun2006,Song2008,Chiang2012,strohmeierIEEEcomsurv}  \\ \midrule

\textbf{ACARS} & \hspace{-0.1bp}\cite{teso2013aircraft,risley2001experimental,defcontruth,smith2016security,yue2015security} & X & X & \hspace{-0.1bp}\cite{risley2001experimental,yue2010approach,ARINC823-1,ARINC823-2,mahmoud2014aeronautical,yue2015security} & \textbf{-}   \\ \midrule

\textbf{TCAS} & \hspace{-0.1bp}\cite{Schaefer13,pierpaoli2015altering} & - & X & \hspace{-0.1bp}\cite{olive2009information,jochum2001encrypted} & \textbf{-}   \\ \midrule

\textbf{GPS} & \hspace{-0.1bp}\cite{Tippenhauer11,humphreys2008assessing,blanch2012satellite}  & - & X & \hspace{-0.1bp}\cite{dovis2015gnss} & \hspace{-0.1bp}\cite{jafarnia2012gps,Tippenhauer11,humphreys2008assessing,lee2015gps,dovis2015gnss,blanch2012satellite}  \\ \midrule

\textbf{ILS} & \hspace{-0.1bp}\cite{middleton2012risk} & - & X & \textbf{-} & \hspace{-0.1bp}\cite{adamy2001ew,adamy2004ew,adamy2008ew,middleton2012risk}  \\ \midrule

\textbf{VOR / NDB / DME} & \hspace{-0.1bp}\cite{adamy2001ew,adamy2004ew,adamy2008ew} & - & X &  \textbf{-} & \hspace{-0.1bp}\cite{adamy2001ew,adamy2004ew,adamy2008ew} \\ \bottomrule
\end{tabular}
\caption{Overview of attacks, security requirements, and existing research on securing wireless aviation communication systems.}
\label{fig:countermeasures-table}
\end{table*}

Comparing the scenarios, we find that the impact of a single ghost aircraft on a radar screen (\textbf{Q1}) is rated as mostly ``minor''. Controllers say it can cause delays during busy times due to additional work and increased separation requirements. It is generally common to experience non-existing radar targets without malicious intent, caused by transponder issues, reflections, clutter, and other reasons. The impact of a ghost aircraft appearing as an intruder on TCAS screens (\textbf{Q2}) is rated slightly higher, it is seen as likely that it would trigger unnecessary manoeuvres that could lead to further unpredictable complications. For both ground radar screens and pilots' TCAS displays, wrong label indications are rated higher in impact compared to ghost aircraft (\textbf{Q3+4}). For both scenarios, more than 50\% consistently rate it as at least a major loss of situational awareness. While controllers and pilots might clarify wrong labels via Voice, the participants do rely on their TCAS data. Selectively missing information is also rated highly in terms of loss of situational awareness (\textbf{Q5+6}). Some comments note that it is inherently difficult or impossible for a pilot or controller to cross-check missing aircraft as they would not even know about it in many situations, thus no action happens. For TCAS, it would lose all purpose, and could have severe impact in bad weather and visibility conditions. Under non-selective jamming attacks on ATC (\textbf{Q7}), operations would be stopped for general aviation and commercial starts and landings would be reduced. All operations would go procedural and use VHF for communication and separation (this is not hypothetical as two major incidents last year in Central Europe proved \cite{Reuters}). For TCAS (\textbf{Q8}), the scenario is rated very similarly to selectively missing information.

For the last scenario (\textbf{Q9}), we asked for the general effect in case the instruments of an aircraft show its position to be different from the true one (e.g., in case of a GPS malfunction, whether caused deliberately or not). If an incorrect position of an aircraft is shown to its pilot, major loss of situational awareness would occur but the specific outcomes are again impossible to predict. The respondents suggest a wide range of scenarios, from additional work for the controller up to a fighter jet escort. One comment noted that German regulations for controllers state that they should not consider outages or wrong labels as a possibility in their work but always rely on the systems (the implication being that otherwise the workload could not be handled), illustrating technology's importance.

\section{Directions in Aviation Security Research}\label{sub:countermeasures}

In this section, we discuss and systematize the work that has been conducted on the security of the analyzed technologies. Besides the works discussed in Section \ref{sub:protocols}, we list in Table \ref{fig:countermeasures-table} the vulnerabilities and defense approaches that have been subject to research in the academic community. In addition, we broadly systematize the existing research on countermeasures into two categories:

\begin{itemize}
\item \textbf{Long-term research of secure new technologies:} These works focus on developing secure new technologies and protocols to be used in aviation communication. Considering the decade-long development, certification, and deployment cycles in aviation, this type of research will see applications only in the long run. The common factor for all such approaches is that they require changes to current technologies or aircraft/ground station equipage which renders them unusable in the short and even medium term.   However, proposals for the ADS-B protocol \cite{wesson2014can,strohmeierIEEEcomsurv,lee2015ads} or TCAS \cite{olive2009information,jochum2001encrypted} for instance show potential directions for future technologies to include security by design, typically using cryptography to ensure integrity and/or confidentiality.  

\item \textbf{Defenses applicable in the short-term:} These works rely on other defense strategies that do not require modifications of existing infrastructure and protocols. Typical representatives of this type of research are separate intrusion detection mechanisms based on cyber-physical defenses such as improved localization protocols (e.g., \cite{Sampigethaya2011}). Also in this category, we can find works on statistical analysis \cite{strohmeier2015lightweight}, machine learning \cite{stelkens2015towards}, physical-layer security \cite{ghose2015verifying,schafer2015secure,schafer16}, or data fusion with backup technologies \cite{monteiro2015detecting2}. The characteristic shared by these approaches is the fact that they can be deployed within a small time window. They would work transparently for aviation users without a costly and time-consuming overhaul of existing systems.
\end{itemize}

As both time horizons are crucial to ensure the continued safety of all services utilizing aviation communication technologies, we note the absence of some research directions in Table \ref{fig:countermeasures-table}. On both the short-term and the long-term side, there has been a plethora of research on the relatively newer ADS-B and its derivative technologies as well as GPS and SSR. There are no existing academic works, which look at securing the legacy technologies PSR and the navigational aids VOR/NDB/DME in the longer term. This can party be attributed to their lower attack surface and the fact that these technologies are being phased out in the foreseeable (yet with 10-20 years fairly distant) future. The same is not true for ILS at many major airports, which will be in use for a very long time, despite the increasing popularity of GPS-based approach systems. Thus, the lack of academic security research on ILS is almost as concerning as it is for VHF, which is also lacking a successor technology at the present, despite its many obvious and documented real-world security challenges.

Furthermore, some technologies lack research on near-term countermeasures that are deployable quickly and without incurring prohibitive costs. This is most relevant for both data link options, ACARS and CPDLC, where all works propose cryptographic measures, with many yet unsolved problems concerning e.g. available packet size or key distribution and management in a globalized and heterogeneous aviation environment \cite{risley2001experimental,yue2010approach,ARINC823-1,ARINC823-2,mahmoud2014aeronautical,yue2015security, mcparland2001securing,patel2015icao,storck2013benefits,getachew2005elliptic,olive2001efficient}. Considering the vulnerabilities discussed in Section \ref{sub:protocols}, this feels like a glaring oversight. Similarly, TCAS is one of the most safety-critical technologies, yet there is no push to improve the immediate security of the system.

Going forward, we believe that in the short-term \textit{awareness} of the issues at hand is a crucial factor. The necessary research can only happen with increased awareness of the aviation system's vulnerabilities, which would motivate the responsible bodies to address the problem. One survey comment noted that regulations are crucial in an industry such as aviation which is very cost-conscious and that actions were typically taken only when required by regulators. Tying this in with the point about awareness, it is clear that the authorities need to be put in a position to issue the necessary regulations required for the deployment of short-term security improvements and long-term secure protocols. An example would be the ACARS protocol, where proprietary cryptography is already used sparsely for parts of the data link communication. A similar move by more airlines could improve the security of this technology quickly.

In terms of research challenges, it is crucial that future security research does not ignore the domain-specific knowledge and requirements of aviation. It is futile to create long-term solutions for protocols that will be replaced such as VHF or SSR. Also, focusing on isolated problems without considering whole system  will inevitably lead to impractical solutions dismissed by the aviation community. Yet, it is inevitable that as aviation requires newly developed secure solutions for all applications, the security community must be involved.   

Some of the aviation experts  noted a lack of appreciation, ignorance and complacency surrounding security within their community. One controller concretely reported a lack of action, related to unawareness of regulators and cost-saving pressures in the industry. As one respondent summarized ``\textit{These questions are silly. Remember aviation is behind 30 years.}'' 

%
\section{Conclusion}\label{sub:Conclusion}
In this work, we provided a systematic analysis of the wireless technologies in aviation. Until now, security analyses of air traffic communication have focused on isolated protocols and did not consider the aviation systems perspective and crucial domain knowledge. Capitalizing on the security knowledge from the academic and hacker communities, technology standards, and the opinions of international aviation experts, we provided a detailed overview of the technologies and their vulnerabilities, existing attacks, and potential countermeasures. 

We further examine the awareness of the aviation community concerning the security of wireless systems and collect expert opinions on the safety impact of attacks on these technologies. Our results motivate the need to reassess the risk of attacks under realistic system models and the development of appropriate countermeasures for the short \textit{and} long term. With our work, we make a first step towards the integration of the wireless security community and the aviation community. A systematic awareness of the existing issues is maybe the most important factor contributing towards safer skies in the future. With the trend going towards more automated data networks communication, we strongly believe that aviation should catch up with the state of the art in wireless security to maintain its excellent safety record in the future. 




\balance
\bibliographystyle{IEEEtran}
\bibliography{literature}
\vspace{-25pt}
\begin{IEEEbiography}[{\includegraphics[width=1in,height=1.25in,clip,keepaspectratio]{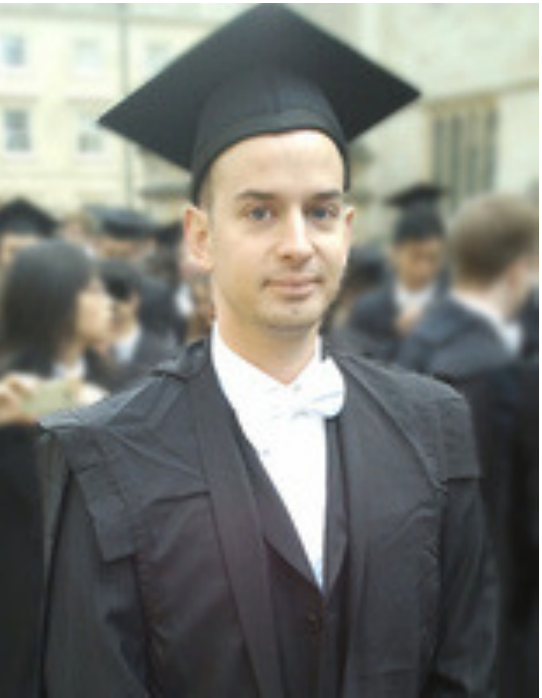}}]
{Martin Strohmeier} (martin.strohmeier @cs.ox.ac.uk) is a DPhil candidate and
teaching assistant in the Department of Computer Science, University of
Oxford. His current research interests are mostly in the area of wireless network
security and air traffic communication Before coming to Oxford, he received his MSc degree from TU Kaiserslautern, Germany and worked as a researcher at Lancaster University's InfoLab21 and Deutsche Lufthansa AG.
\end{IEEEbiography}
 \vspace{-25pt}

 \begin{IEEEbiography}[{\includegraphics[width=1in,height=1.25in,clip,keepaspectratio]{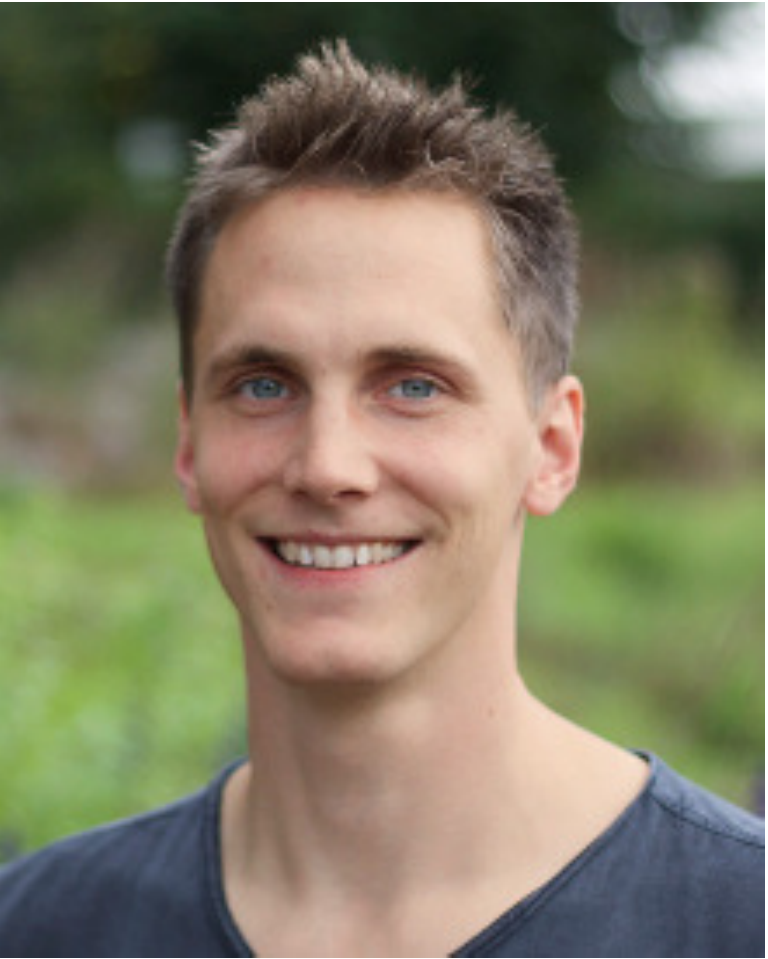}}]
{Matthias Schäfer} (schaefer@cs.uni-kl.de) is a PhD candidate in the Department of Computer Science at the University of Kaiserslautern, Germany, where he also received his MSc degree in computer science in 2013. Between 2011 and 2013, he worked for the Information Technology and Cyberspace group of armasuisse, Switzerland and visited the Department of Computer Science of the University of Oxford, UK, as a visiting researcher. He is a co-founder and board member of the OpenSky Network association.
 \end{IEEEbiography}
\vspace{-25pt}

\begin{IEEEbiography}
[{\includegraphics[width=1in,height=1.25in,clip,keepaspectratio]{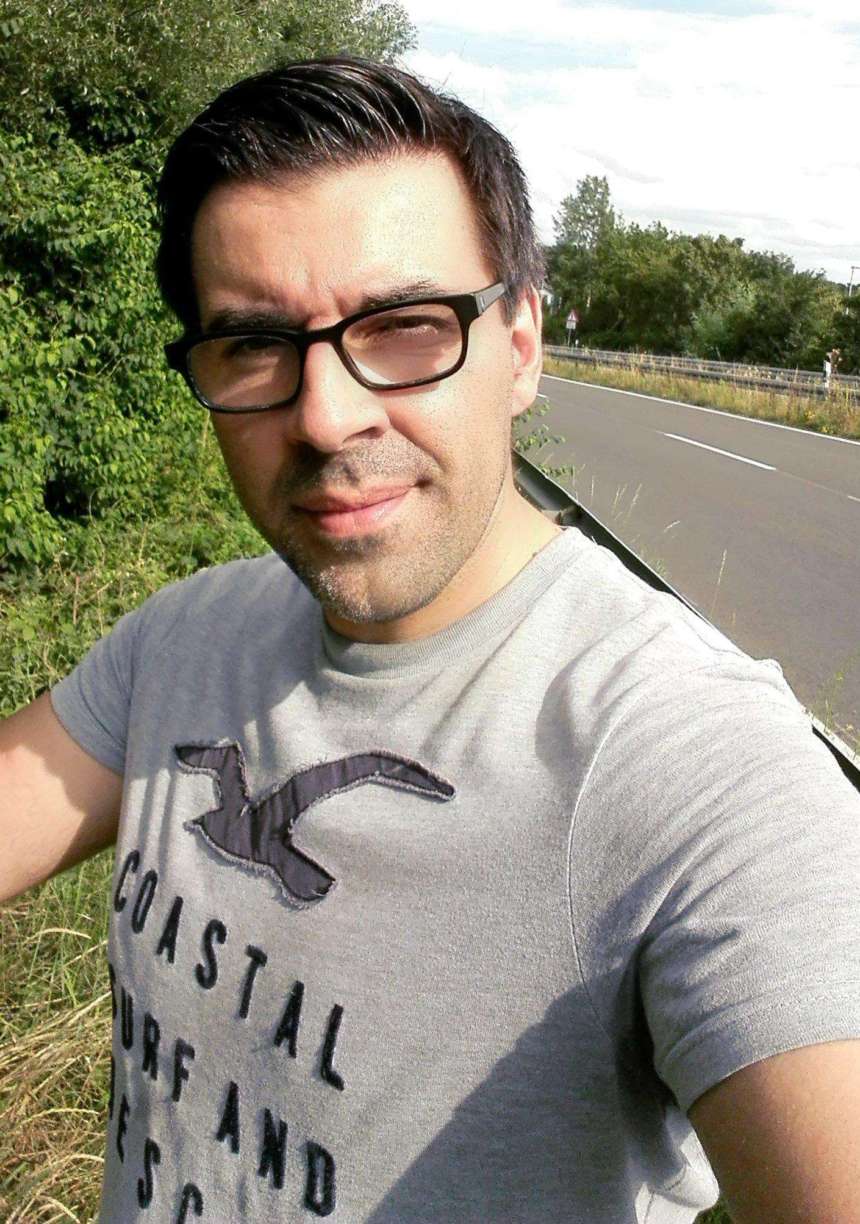}}]
{Rui Pinheiro} (pinheiro@opensky-network.org) is an air traffic controller with 15 years of professional experience at a major Area Control Center in Germany. He is currently studying for a Computer Science degree at University of Hagen. Besides his aviation expertise, he is highly proficient in air traffic communication technology and holds a ham radio licence.
 \end{IEEEbiography}
 \vspace{-25pt}

\begin{IEEEbiography}[{\includegraphics[width=1in,height=1.25in,clip,keepaspectratio]{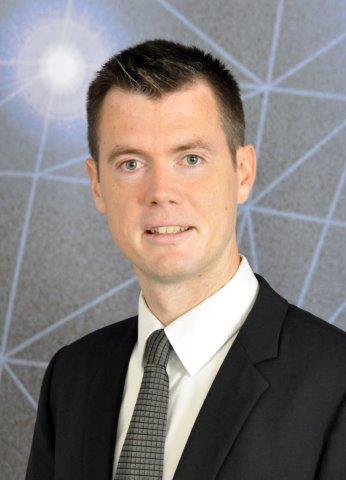}}]
{Vincent Lenders} (vincent.lenders@armasuisse.ch) is a research program manager at armasuisse heading the cyberspace and information research activities of the Swiss Armed Forces. He received his MSc and PhD in electrical engineering from ETH Zurich, Switzerland in 2001 and 2006, respectively, and was a postdoctoral researcher at Princeton University, USA in 2007. He serves as industrial director of the Zurich Information Security and Privacy Center at ETH Zurich and is co-founder and board member of the OpenSky Network and Electrosense associations.
\end{IEEEbiography}
\vspace{-25pt}

\begin{IEEEbiography}[{\includegraphics[width=1in,height=1.25in,clip,keepaspectratio]{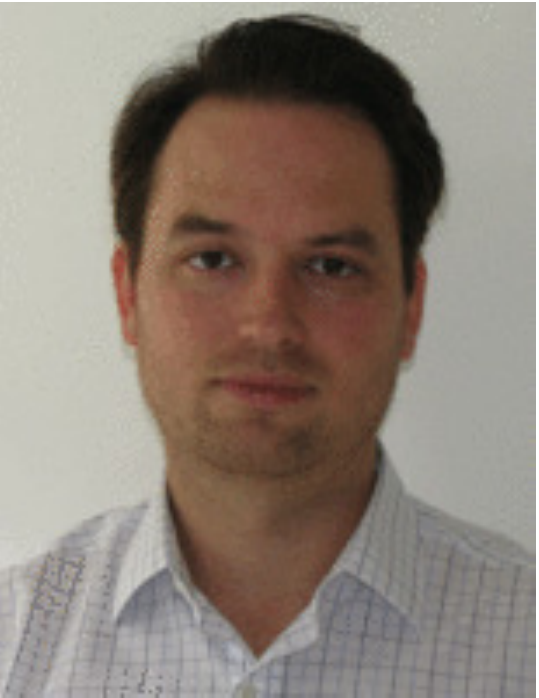}}]
{Ivan Martinovic} (ivan.martinovic@cs.ox.ac.uk) is an Associate Professor at the Department of Computer Science, University of Oxford. Before coming to Oxford, he was a postdoctoral researcher at the Security Research Lab, UC Berkeley and the Secure Computing and Networking Centre, UC Irvine. He obtained his PhD from TU Kaiserslautern and MSc from TU Darmstadt, Germany.
\end{IEEEbiography}

\end{document}